\documentclass[11pt]{article}
\usepackage{amsmath, amssymb, graphicx, dsfont, xcolor, natbib, atbegshi, amsthm}
\usepackage[margin= 1in]{geometry}

\definecolor{darkgreen}{rgb}{0.0, 0.2, 0.13}
\definecolor{ao}{rgb}{0.0, 0.5, 0.0}
\definecolor{blush}{rgb}{0.87, 0.36, 0.51}

\newtheorem{theorem}{Theorem}

\newtheorem{lemma}{Lemma}

\begin{document}

\title{Nonparametric inference on non-negative dissimilarity measures at the boundary of the parameter space}
\author{Aaron Hudson \\ Fred Hutchinson Cancer Center}
\date{}
\maketitle
\bibliographystyle{biom}

\begin{abstract}
It is often of interest to assess whether a function-valued statistical parameter, such as a density function or a mean regression function, is equal to any function in a class of candidate null parameters. This can be framed as a statistical inference problem where the target estimand is a scalar measure of dissimilarity between the true function-valued parameter and the closest function among all candidate null values. These estimands are typically defined to be zero when the null holds and positive otherwise.  While there is well-established theory and methodology for performing efficient inference when one assumes a parametric model for the function-valued parameter, methods for inference in the nonparametric setting are limited. When the null holds, and the target estimand resides at the boundary of the parameter space, existing nonparametric estimators either achieve a non-standard limiting distribution or a sub-optimal convergence rate, making inference challenging. In this work, we propose a strategy for constructing nonparametric estimators with improved asymptotic performance.  Notably, our estimators converge at the parametric rate at the boundary of the parameter space and also achieve a tractable null limiting distribution. As illustrations, we discuss how this framework can be applied to perform inference in nonparametric regression problems, and also to perform nonparametric assessment of stochastic dependence.
\end{abstract}

\newpage

\section{Introduction}

Suppose we are interested in studying a function-valued parameter of an unknown probability distribution, such as a conditional mean function or a density function. 
For such parameters, one can typically define a goodness-of-fit functional, which measures the closeness of any given candidate function to the true population parameter.
The goodness-of-fit achieves its minimum when evaluated at the true population parameter.  
It is often of scientific interest to compare multiple models for the function-valued parameter.
In particular, one may seek to determine whether the minimizer of the goodness-of-fit over a, possibly large, function class is equal to the minimizer over a smaller sub-class.
The difference between the minima over reduced and full function classes can serve as a natural measure of dissimilarity for comparing the corresponding minimizers.
This dissimilarity measure is non-negative, with values of zero corresponding to no dissimilarity.
The main focus of this work is on estimation of such dissimilarity measures and testing the null hypothesis of no dissimilarity, or equality of goodness-of-fit.

As an example, suppose that an investigator would like to determine whether an exposure is conditionally associated with an outcome,  given a set of confounding variables.
This can be formulated as a statistical inference problem, where the objective is to determine whether the conditional mean of the outcome, given both the exposure and confounders, is equivalent to the conditional mean of the outcome, given only the confounders.
One can specify a full model for the conditional mean as a class of functions that depends on both the exposure and confounders, while the reduced model is the subclass of functions that may depend on the confounders but do not depend on the exposure.
Several goodness-of-fit measures, such as the expected squares error loss, can be used to assess how close a candidate parameter is to the conditional mean given the exposure and confounders.
And so, one can test for conditional independence by assessing whether the best approximation of the conditional mean in the full model class is an improvement over the best approximation in the reduced class, in terms of the goodness-of-fit.

When the function-valued parameter of interest is modeled using a finite-dimensional function class, there are standard procedures available for performing inference.  
For instance, the classical likelihood ratio test is widely-used to compare classes of regression functions when the conditional distribution of the outcome given the predictor and covariates is assumed to belong to a parametric family of probability distributions \citep{wilks1938large}.
There also exist approaches for efficient inference in settings where the reduced and full function classes are both infinite-dimensional, but the difference between the two classes is finite dimensional.
For instance, in regression problems of the form described in the example above, it is common to assume that the conditional mean of the outcome given the exposure and covariates follows a partially linear model.
In a partially linear model, the full conditional mean can be expressed as the sum of an unknown function of the confounders, which is only assumed to belong to a large infinite-dimensional function class, plus a linear function of the exposure of interest.
One can therefore assess for conditional dependence by determining whether the linear function has zero slope, which is a well-studied inference problem \citep{chernozhukov2018double, bhattacharya1997semiparametric, robinson1988root, donald1994series}.

In this work, we focus on the more challenging setting in which the difference between the full and reduced function classes is infinite-dimensional.
Recently, several investigators have examined whether modern methods for estimation of smooth functionals of unknown probability distributions in a nonparametric model, such as targeted-minimum loss-based estimation \citep{van2011targeted, van2018targeted} and one-step estimation \citep{pfanzagl1982contributions}, can be applied to attain inference on non-negative dissimilarity measures \citep{williamson2021nonparametric, williamson2021general, hines2022variable, kennedy2023semiparametric, kandasamy2015nonparametric}.
For these estimation strategies to be viable, the target estimand -- in this case the non-negative dissimilarity measure -- must be a pathwise differentiable functional of the underlying probability distribution with non-zero pathwise derivative.
In essence, this means that the target estimand makes smooth but non-negligible changes in response to infinitesimally small perturbations around the unknown probability distribution.
While pathwise differentiability of the target can be established in many examples, the pathwise derivative is typically zero when the null hypothesis of no dissimilarity holds.
That the derivative is zero can be seen as a consequence of the fact that, under the null, the target estimand achieves its minimum at the true unknown distribution.
In this setting, conventional estimation strategies do not achieve parametric-rate convergence or attain tractable limiting distributions, making hypothesis testing challenging.

When the target estimand satisfies additional smoothness assumptions, it can be possible to construct estimators with improved asymptotic behavior by utilizing higher-order pathwise derivatives \citep{pfanzagl1985asymptotic, robins2008higher, van2014higher, carone2018higher}.
While this approach has been successful in some examples \citep{luedtke2019omnibus}, it is seemingly rare that for a given statistical functional, higher-order pathwise derivatives exist, so this strategy does not appear to be broadly applicable.

In this work, we propose a general method for estimation and inference on non-negative dissimilarity measures.
Our proposal builds upon recent developments on the construction of omnibus tests for equality of function-valued parameters to fixed null parameters \citep{hudson2021inference, westling2021nonparametric}.
The key idea used is that one can perform inference on a function-valued parameter by estimating a large collection of simpler one-dimensional estimands that act as an effective summary thereof.
Here, we show that in many instances, non-negative dissimilarity measures can be represented as the largest value in a collection of simple one-dimensional estimands.
In such cases, we can estimate non-negative dissimilarity measures using the maximum of suitably well-behaved estimators for these scalar quantities. 
Our main results show that when efficient estimators for the simple estimands are used, the resulting estimator for the non-negative dissimilarity measure achieves parametric-rate convergence under the null and also attains a tractable limiting distribution.
This makes it possible to construct well-calibrated asymptotic tests of the null.
We also show that when the alternative holds, our estimator is asymptotically efficient.
To the best of our knowledge, our work is the first to provide a general theoretical basis for recovering parametric rate inference on non-negative dissimilarity measures in a nonparametric model.

The remainder of the paper is organized as follows.
In Section \ref{sec:prelim}, we formally introduce the class of non-negative dissimilarity measures of interest, and we describe some motivating examples.
In Section \ref{sec:plug-in}, we review an existing approach for inference based on plug-in estimation and provide a discussion of some of its limitations.
In Section \ref{sec:framework}, we propose a new estimator for non-negative dissimilarity measures, and we describe its theoretical properties.
In Section \ref{sec:bootstrap}, we present multiplier bootstrap methods for testing the null of no dissimilarity, and for constructing confidence intervals.
In Section \ref{sec:implementation} we discuss implementation and practical concerns.
In Section \ref{sec:illustration1}, we illustrate how our methodology can be used to perform inference in a nonparametric regression model.
We present results from our simulation study in Section \ref{sec:simulations}, and we conclude with a discussion in Section \ref{sec:discussion}.

\section{Preliminaries} \label{sec:prelim}

\subsection{Data structure and target estimand} 

Let $Z_1, \ldots, Z_n$ be \textit{i.i.d.} random vectors, generated from an unknown probability distribution $P_0$.
We make few assumptions about $P_0$ and only require that it belongs to a flexible nonparametric model $\mathcal{M}$, which is essentially unrestricted, aside from mild regularity conditions.
For a given probability distribution $P$ in $\mathcal{M}$, let $\theta_P$ be a function-valued summary of interest with domain $\mathcal{O} \subseteq \mathbb{R}^d$ for a positive integer $d$ and range $\mathcal{K} \subseteq \mathbb{R}$.
We denote by $\theta_{0} := \theta_{P_0}$ the evaluation of this summary at $P_0$.

Suppose that $\theta_P$ is known to belong to a, possibly infinite-dimensional, function class $\Theta$.
For a given distribution $P$, we define a real-valued functional $G_P: \Theta \to \mathbb{R}$ that satisfies
\begin{align}
G_P(\theta_P) = \inf_{\theta \in \Theta} G_P(\theta). \label{best-fit}
\end{align}
The functional $G_P$ measures the goodness-of-fit of any function $\theta \in \Theta$ -- larger values of $G_P(\theta)$ indicate that $\theta$ and $\theta_P$ are farther away from one another, in a sense. 
Throughout this paper, we use the shorthand notation $G_0 := G_{P_0}$ to denote the value of the goodness-of-fit measure at $P_0$.

Let $\Theta^* \subset \Theta$ be a subclass of $\Theta$, and let $\theta_P^*$ be a function that satisfies
\begin{align*}
G_P(\theta_P^*) = \inf_{\theta \in \Theta^*} G_P(\theta).
\end{align*}
In essence, $\theta^*_P$ is the closest function to $\theta_P$ among all functions in the subclass $\Theta^*$.
We define as our target parameter the difference between the goodness-of-fit of $\theta_P$ and $\theta^*_P$, 
\begin{align}
\Psi_P := G_P(\theta^*_P) - G_P(\theta_P),
\label{improvement in fit}
\end{align}
and we again use the shorthand notation $\Psi_0 := \Psi_{P_0}$.
Throughout this manuscript, we refer to $\Psi_0$ as the \textit{improvement in fit} because it represents the improvement in the goodness-of-fit attained by using the full function class instead of the reduced class.

Because $\Theta^*$ is contained within $\Theta$, it can be seen that $\Psi_P$ is a non-negative statistical functional, and $\Psi_P$ is only equal to zero when $\theta_P$ provides no improvement in fit compared with $\theta^*_P$.
In many applications, a problem of central importance is to determine whether $\theta^*_P$ is inferior to $\theta_P$ in terms of goodness-of-fit.
Letting $\Psi_{0} := \Psi_{P_0}$, we are interested in performing a test of the null hypothesis
\begin{align}
H_0: \Psi_0 = 0.
\label{null}
\end{align}
Additionally, because statistical functionals that have the representation in \eqref{improvement in fit} have scientifically meaningful interpretations in some contexts,  estimation of $\Psi_0$ and confidence interval construction are also of practical interest.
Our paper provides a general framework for estimation, testing, and confidence interval construction for statistical functionals of this form.

\subsection{Examples} \label{sec:examples}

In what follows, we introduce some working examples.
As a first example, we discuss statistical inference in nonparametric regression models, and second, we discuss a nonparametric approach for assessing dependence between a pair of random variables.
We then describe a simple way to define a goodness-of-fit measure for any function-valued parameter.

\noindent \textbf{Example 1: Inference in a Nonparametric Regression Model}
\\
Let $Z = (W, X, Y)$, where $Y \in \mathbb{R}$ is a real-valued outcome variable, and $X \in \mathbb{R}^{d_1}$ and $W \in \mathbb{R}^{d_2}$ are vectors of predictor variables with dimensions $d_1$ and $d_2$, respectively.
We define $\Theta$ as a (possibly large) class of prediction functions with domain $\mathbb{R}^{d_1 + d_2}$ and range $\mathbb{R}$.
Each function $\theta \in \Theta$ takes as input a realization $(w, x)$ of the predictor vector $(W, X)$ and returns as output a predicted outcome.

We are interested in studying the conditional mean of the outcome given the predictors, defined as $\theta_P: (w,x) \mapsto E_P[Y|X = x, W = w]$.
It is well-known that the conditional mean can be characterized as the minimizer of the expected squared error loss over $\Theta$, if $\Theta$ is sufficiently large.
That is, defining the goodness-of-fit measure
\begin{align*}
G_P: \theta \mapsto \int \left\{y - \theta(w,x)\right\}^2 dP(w,x,y), 
\end{align*}
the conditional mean satisfies $G_P(\theta_P) = \inf_{\theta \in \Theta} G_P(\theta)$.

Consider now the set of candidate prediction functions that do not depend on $X$, which we write as
\begin{align*}
\Theta^* := \left\{\theta \in \Theta: \theta(w,x_1) = \theta(w, x_2) \text{ for every } x_1 \neq x_2 \right\}.
\end{align*}
When $\Theta$ is large, any minimizer $\theta^*_P$ of the expected squared error loss over $\Theta^*$ is almost everywhere equal to the conditional mean of $Y$ given $W$.
We are often interested in determining whether $X$ is an important set of predictors in the sense that it does not need to be included in a prediction function in order for optimal squared error loss to be achieved.
If $X$ is not important in this sense, the conditional mean of $Y$ given $X$ and $W$ does not depend on $X$, and the difference in the expected squared error loss $\Psi_0 = G_0(\theta^*_0) - G_0(\theta_0)$ is zero.
Otherwise, $\Psi_0$ is positive.
Thus, assessing variable importance can be framed a statistical inference problem of the type described in Section \ref{sec:prelim}.

Many recent works have studied inference on variable importance estimands of a similar form to that we describe above \citep[see, e.g.,][]{williamson2021nonparametric, williamson2021general, verdinelli2021decorrelated, zhang2020floodgate}.
These works all encounter difficulties with constructing estimators for their original target estimand that achieve parametric rate convergence under the null.
To the best of our knowledge, there is currently no solution available to this problem.

%
%
%
%
%

\noindent \textbf{Example 2: Nonparametric Assessment of Stochastic Dependence}
\\
Let $Z = (X, Y)$, where $X \in \mathbb{R}$ and $Y \in \mathbb{R}$ are real-valued random variables, and let $\theta_P$ denote the log of the joint density of $(X,Y)$ under $P$ with respect to some dominating measure $\nu$.
Our objective here is to determine whether $X$ and $Y$ are dependent.
If $X$ and $Y$ are independent, by basic laws of probability, the joint density function can be expressed as the product of the marginal density functions, i.e.,
\begin{align*}
\exp \theta_P(x,y) = \int \exp \theta_P(x,y_1) \nu (dy_1) \int \exp \theta_P(x_1,y) \nu(dx_1)
\end{align*}
for all $x, y \in \mathbb{R}$.
We can therefore assess dependence between $X$ and $Y$ by defining a goodness-of-fit measure for the joint density function, and determining whether the goodness-of-fit of the true joint density is lower than the goodness-of-fit of the product of the marginal densities.

Let $\Theta$ be a collection of candidate values for the log density function, and assume that $\Theta$ is large enough to contain $\theta_0$, the log density under $P_0$.
The density function can be represented as a minimizer of the expected cross-entropy loss.
Therefore, defining the goodness-of-fit measure
\begin{align*}
G_P: \theta \mapsto -\int \theta(x,y) dP(x,y), 
\end{align*}
the joint density satisfies \eqref{best-fit}.

We now define $\Theta^*$ as the class of candidate log density functions for which the joint density can be expressed as the product of two marginal density functions -- that is,
\begin{align*}
\Theta^* := \left\{ \theta \in \Theta: \exp \theta(x,y) = \int \exp \theta(x,y_1) \nu (dy_1) \int \exp \theta(x_1,y) \nu(dx_1) \text{ for all } x,y \in \mathbb{R} \right\}.
\end{align*}
Any minimizer $\theta^*_P$ of $G_P$ over $\Theta^*$ is almost everywhere equal to the product of the marginal densities of $X$ and $Y$ under $P$.
Therefore, $\Psi_0 := G_0(\theta^*_0) - G_0(\theta_0)$ is zero if $X$ and $Y$ are independent, and $\Psi_0$ is otherwise positive.
One can assess dependence between $X$ and $Y$ by performing inference on $\Psi_0$, so similar to the previous example, this problem falls within our framework.

The measure of dependence $\Psi_0$ we have defined here is commonly referred to as the \textit{mutual information} and has been a widely-studied measure of stochastic dependence \citep[see, e.g., ][]{paninski2003estimation, steuer2002mutual}. 
We are not aware of an existing nonparametric estimator for the mutual information that achieves parametric rate convergence under the null of independence.
This appears to be a longstanding open problem.

\noindent \textbf{Example 3: Generic $L_2$ Distance}

Suppose one is interested in assessing whether a given function-valued parameter $\theta_P$ is equal to a fixed and known function $\theta^*$.
For a measure $\nu$ on $\mathcal{O}$, one can define as a goodness-of-fit measure an integrated squared difference between $\theta_P$ and $\theta^*$:
\begin{align*}
G_P: \theta \mapsto \int \left\{ \theta(o) - \theta_P(o) \right\}^2 d\nu(o).
\end{align*}
Because $G_P$ is non-negative, and $G_P(\theta_P) = 0$, it is easy to see that $G_P$ is minimized by $\theta_P$.

One might wish to perform inference on the quantity 
\begin{align*}
\Psi_0 = G(\theta^*) - G_0(\theta_0) = \int \left\{ \theta^*(o) - \theta_0(o) \right\}^2 d\nu(o).
\end{align*}
Clearly, $\Psi_0$ is equal to zero only when $\theta_0$ is equal to $\theta^*$ almost everywhere $\nu$.
Estimands of this form can be of interest when one wishes to construct an omnibus test of the hypothesis that $\theta_0 = \theta^*$.
The framework we develop in this paper can be applied in this setting as well, and so, methodology for inference on the improvement in fit can be seen as generally useful for performing inference on function-valued parameters.

\section{Plug-in estimation of the improvement in fit} \label{sec:plug-in}



We now describe an approach for nonparametric inference on $\Psi_0$ based on plug-in estimation, and we discuss the shortcomings of this approach.
The methodology we describe below and its limitations are discussed extensively by \cite{williamson2021general} in the context of nonparametric regression, though their theoretical and methodological results are more broadly applicable.

Suppose that for any $\theta \in \Theta$, $G_P(\theta)$ is a \textit{pathwise differentiable} functional of $P$, meaning that $G_P(\theta)$ changes smoothly with respect to small changes in $P$ \citep{bickel1998efficient}. 
When $G_P(\theta)$ is pathwise differentiable, it is generally possible to construct an estimator $G_n(\theta)$ that is asymptotically linear in the sense that
\begin{align}
G_n(\theta) - G_0(\theta) = \frac{1}{n} \sum_{i=1}^n \phi_{P_0}(Z_i; \theta) + r_n(\theta),
\label{asymp-lin-risk}
\end{align}
where $\phi_{P_0}(Z;\theta)$ has mean zero and finite variance under $P_0$, and $r_n(\theta) = o_P(n^{-1/2})$ is an asymptotically negligible remainder term.
The function $\phi_{P_0}(\cdot;\theta)$ determines the first order asymptotic behavior of $G_n(\theta)$ and is commonly referred to as the \textit{influence function} of $G_n(\theta)$.
Because $G_n(\theta)$ is asymptotically linear, it is $n^{1/2}$-rate consistent and asymptotically Gaussian by the central limit theorem.
Conventional strategies for constructing asymptotically linear estimators include one-step estimation \citep{pfanzagl1982contributions} and targeted minimum loss-based estimation \citep{van2011targeted, van2018targeted}.

Given an asymptotically linear estimator $G_n$ for $G_0$, we can obtain estimators $\theta_n$ and $\theta_n^*$  for $\theta_0$ and $\theta^*_0$ by minimizing $G_n$ over $\Theta$ and $\Theta^*$, respectively.
That is, we take
\begin{align*}
\theta_n := \underset{\theta \in \Theta}{\text{arg min}} \, G_n(\theta), \quad
\theta^*_n := \underset{\theta \in \Theta^*}{\text{arg min}} \, G_n(\theta).
\end{align*}
We can then obtain the following plug-in estimator $\Psi_n$ for $\Psi_0$:
\begin{align*}
\Psi_n := G_n(\theta^*_n) - G_n(\theta_n).
\end{align*}
It can be shown that, under mild regularity conditions, the plug-in estimator is asymptotically linear with influence function $\phi_{P_0}(\cdot;\theta_0) - \phi_{P_0}(\cdot; \theta^*_0)$ \citep{williamson2021general}. That is, the plug-in estimator satisfies 
\begin{align}
\Psi_n - \Psi_0 = \frac{1}{n}\sum_{i=1}^n\phi_{P_0}(Z_i;\theta^*_{0}) - \phi_{P_0}(Z_i;\theta_0) + o_P(n^{-1/2}).
\label{eqn:plugin-exp}
\end{align}
From an initial inspection, it would appear that there is no loss in efficiency resulting from estimating $\theta_0$ and $\theta_0^*$.
That is, if $\theta_0$ and $\theta_0^*$ were known, then the estimator $G_n(\theta_0) - G_n(\theta^*_0)$ would have the same asymptotically linear representation as $\Psi_n$.

Under the null, $G_n(\theta_n)$ and $G_n(\theta^*_n)$ have the same influence function, and the leading term in \eqref{eqn:plugin-exp} vanishes.  
Therefore, the convergence rate and limiting distribution of the plug-in estimator are determined by the higher-order remainder term. 
When $\Theta$ is a finite-dimensional model, it is often possible to establish that, under the null, the remainder term is $O_P(n^{-1})$ and attains a tractable limiting distribution.
Conversely, in the infinite-dimensional setting, the remainder typically converges at a slower-than-$n$ rate, and its asymptotic distribution is difficult to characterize.
This makes it challenging to approximate the null sampling distribution of $\tilde{\Psi}_n$ and hence challenging to construct a hypothesis test for no improvement in fit.
Moreover, confidence intervals based on a normal approximation to the sampling distribution can fail to achieve the nominal coverage rate when $\Psi_0 = 0$.



In order to develop an estimator for $\Psi_0$ that has better asymptotic properties than the plug-in, it is helpful for us to further investigate what is the source of the plug-in estimator's poor behavior.
We can first recognize that, as $\Psi_0$ is a measure of an improvement in fit, estimating $\Psi_0$ involves performing a search away from $\theta^*_0$ to identify whether any candidate function in the difference between the full and reduced function classes, $\Theta \setminus \Theta^*$, provides a better fit than $\theta^*_0$.

Suppose now that $\Theta \setminus \Theta^*$ can be expressed as a collection of, potentially many, one-dimensional sub-models.
Let $g$ be a fixed function from $\mathcal{K} \times \mathbb{R}$ to $\mathcal{K}$ that satisfies $g(k; 0) = k$ for any $k \in \mathcal{K}$.
For a scalar $\beta$ and a fixed function $f: \mathcal{O} \to \mathbb{R}$, we define $\theta^*_{P,f}$ as the one-dimensional sub-model
\begin{align}
\theta_{P,f}^*(\cdot; \beta): o \mapsto g(\theta^*_{P}(o), \beta f(o)).
\label{sub-model}
\end{align}
We have constructed our sub-model $\theta^*_{P,f}$ so that it passes through the null best fit $\theta^*_P$ at $\beta = 0$, i.e.,
\begin{align}
&\theta^*_{P,f}(\cdot; 0) = \theta^*_P(\cdot). \label{submod-prop-1}
\end{align}
We can therefore interpret $f$ as the path along which $\theta^*_{P,f}$ approaches $\theta^*_P$ as $\beta$ tends to zero.
We assume that there exists a function class $\mathcal{F}$ and a symmetric interval $\mathcal{B}$ such that 
\begin{align*}
\Theta \setminus \Theta^* = \left\{\theta = \theta^*_{P,f}(\cdot; \beta): f \in \mathcal{F}, \beta \in \mathcal{B}\right\}.
\end{align*}
We will see that using this representation for our model facilitates making comparisons between any function in $\Theta \setminus \Theta^*$ and the best null fit.

We now define $G_{P,f}: \beta \to \mathbb{R}$ as the goodness-of-fit of $\theta^*_{P,f}(\cdot; \beta)$, i.e.,
\begin{align}
G_{P,f}(\beta) := G_{P}(\theta^*_{P,f}(\cdot; \beta)),
\end{align}
and similarly as above, we use the shorthand notation $G_{0,f} := G_{P_0,f}$.
We assume that $G_{P,f}$ is a smooth convex function, and we denote the first and second derivatives of $G_{P,f}$ in $\beta$ by
\begin{align}
G'_{P,f}(\beta) := \frac{d}{d\beta} G_{P,f}(\beta), \quad
G''_{P,f}(\beta) := \frac{d^2}{d\beta^2} G_{P,f}(\beta).
\end{align}
We define $\beta_{P,f}$ as the minimizer of the goodness-of-fit measure along the parametric sub-model over the interval $\mathcal{B}$:
\begin{align*}
\beta_{P,f} := \underset{\beta \in \mathcal{B}}{\text{arg min}} G_{P,f}(\beta).
\end{align*}
Due to the convexity of $G_{P,f}$, for large enough $\mathcal{B}$, $\beta_{P,f}$ is the unique solver of $G'_{P,f}(\beta_{P,f}) = 0$.
Under this regime, $\theta_P$ satisfies $G_P(\theta_P) = \inf_{f \in \mathcal{F}} G_{P,f}(\beta_{P,f})$, and we can write $\Psi_P$ as
\begin{align*}
\Psi_P = \sup_{f \in \mathcal{F}} G_{P,f}(0) - G_{P,f}(\beta_{P,f}).
\end{align*}
We can see that, in view of condition \eqref{submod-prop-1}, $\Psi_P = 0$ only when $\sup_{f \in \mathcal{F}} |\beta_{P,f}| = 0$.

Let $G_n$ and $\theta^*_n$ be the estimators for $G_0$ and $\theta^*_0$  described in earlier in this section, and let $\theta^*_{n,f}(\cdot;\beta): o \mapsto g(\theta_n^*(o), \beta f(o))$ be the plug-in estimator for the sub-model.
We define the plug-in estimator for $G_{0,f}$ as
\begin{align*}
G_{n,f}(\beta) := G_n(\theta_{n,f}^*(\cdot; \beta)),
\end{align*}
and we write its first and second derivatives as
\begin{align*}
G'_{n,f}(\beta) := \frac{d}{d\beta} G_{n,f}(\beta), \quad G''_{n,f}(\beta) := \frac{d^2}{d\beta^2} G_{n,f}(\beta).
\end{align*}
We define the plug-in estimator $\beta_{n,f}$ for $\beta_{0,f}$ as the minimizer of $G_{n,f}$ over $\mathcal{B}$.
For large $\mathcal{B}$, $\beta_{n,f}$ satisfies $G'_{n,f}(\beta_{n,f}) = 0$ for all $f$ in $\mathcal{F}$, and the plug-in estimator $\theta_n$ for $\theta_{0}$ satisfies
\begin{align*}
G_n(\theta_n) = \inf_{f \in \mathcal{F}} G_{n,f}(\beta_{n,f}).
\end{align*}
The plug-in estimator for $\Psi_0$ can therefore be expressed as
\begin{align*}
\Psi_n = \sup_{f \in \mathcal{F}} G_{n,f}(0) - G_{n,f}(\beta_{n,f}) .
\end{align*}

Using this representation for the plug-in estimator $\Psi_n$ makes it easier for us to carefully study its asymptotic behavior in the setting where $\Psi_0 = 0$.
By performing a second order Taylor expansion for $G_{n,f}$ around $\beta_{n,f}$ for every $f \in \mathcal{F}$, we can write the plug-in as
\begin{align*}
\Psi_{n} = \sup_{f \in \mathcal{F}} G'_{n,f}(\beta_{n,f}) (0-\beta_{n,f}) + \frac{1}{2}G''_{n,f}(\beta_{n,f}) \beta_{n,f}^2 + r_{n},
\end{align*}
where $r_n$ is a higher order remainder term that should approach zero at a faster rate than the leading terms.
Because $G'_{n,f}(\beta_{n,f}) = 0$ for all $f$, the first term in this expansion vanishes, leaving us with
\begin{align*}
\Psi_n  = \sup_{f \in \mathcal{F}} 
\frac{1}{2}G''_{n,f}(\beta_{n,f})\beta_{n,f}^2 + r_n.
\end{align*}
If $G''_{n,f}(\beta_{n,f})$ is consistent for $G''_{0,f}(\beta_{0,f})$ uniformly in $\mathcal{F}$, then by Slutsky's theorem, one can replace the random quantities $G''_{n,f}(\beta_{n,f})$ with the fixed values $G''_{0,f}(\beta_{0,f})$ in the above display.
It would appear then that, under the null, the limiting distribution of $\Psi_n$ is determined by the behavior of the stochastic process $\{\beta_{n,f}: f \in \mathcal{F}\}$.

If it were possible to characterize the joint limiting distribution of $\left\{\beta_{n,f}: f \in \mathcal{F}\right\}$ under the null where $\beta_{0,f} = 0$ for all $f \in \mathcal{F}$, the limiting distribution of $\Psi_n$ could be characterized using a straightforward application of the continuous mapping theorem.
Typically, $\beta_{0,f}$ is a pathwise differentiable parameter for each $f \in \mathcal{F}$, making it possible to construct estimators thereof that converge at a $n^{1/2}$-rate and achieve an Gaussian limiting distribution.
Ideally, one would be able to establish that the standardized process $\left\{ n^{1/2} \left[\beta_{n,f} - \beta_{0,f}\right]: f \in \mathcal{F} \right\}$ converges weakly to a Gaussian process as long as the collection of paths $\mathcal{F}$ is not overly complex.
However, in many settings, this property is not satisfied by the plug-in estimator.
One can view the plug-in estimator $\beta_{n,f}$ for $\beta_{0,f}$ as a functional of the estimator $G_{n,f}$ for $G_{0,f}$.
As stated above, estimating $G_{0,f}$ requires us to estimate the nuisance parameter $\theta^*_0$.
In settings where $\Theta^*$ is a large nonparametric function class, our estimator $\theta^*_n$ for $\theta^*_0$ will necessarily converge slower than the parametric rate of $n^{1/2}$ and may retain non-negligible asymptotic bias.
Consequently, $\theta^*_n$ generates bias for $G_{n,f}$, which leads to $\beta_{n,f}$ retaining non-negligible bias as well.
Indeed, $\beta_{n,f}$ will typically converge slower than the parametric rate of $n^{1/2}$, causing $\Psi_n$ to converge at a sub-optimal rate and achieve a non-standard limiting distribution.

To summarize, estimating $\Psi_0$ requires one to perform a search away from $\theta^*_P$ in order to attempt to identify a candidate function in $\Theta \setminus \Theta^*$ that provides an improvement in the goodness-of-fit.
In the regime we describe above, performing this search is equivalent to finding the best fit along each parametric sub-model that passes through the null, and subsequently taking the best fit among all of the sub-models.
From the above argument, we can see that the plug-in estimator has poor asymptotic properties because the plug-in estimator for the best fit along the parametric sub-models can be sub-optimal when $\Theta^*$ is large.
Thus, the key to obtaining an estimator with improved asymptotic properties is to efficiently estimate the best fit along each of the sub-models that comprise $\Theta \setminus \Theta^*$.

\section{Bias-corrected estimation of the improvement in fit} \label{sec:framework}

From the discussion in Section \ref{sec:plug-in}, it would seem that if one had an efficient estimator for the goodness-of-fit along each parametric sub-model, and hence an efficient estimator for $\beta_{0,f}$, one could obtain an estimator for the improvement in fit $\Psi_0$ that has better asymptotic properties than the plug-in.  In what follows, we describe a general strategy for constructing an estimator that has a tractable limiting distribution when $\Psi_0$ is at the boundary of the parameter space. We show that our newly-proposed estimator enjoys the same $n$-rate convergence that is typically attained in parametric models. 

\subsection{Uniform inference along the parametric sub-models} \label{sec:submodel-inference}

Our proposal requires us to construct an estimator for $\{G_{0,f}(\beta): f\in\mathcal{F}, \beta \in \mathcal{B}\}$ that enables us to perform inference uniformly along the collection of parametric sub-models.
In this sub-section, we first outline a set of sufficient conditions under which an estimator has asymptotic properties that facilitate uniform inference.
We then describe a strategy for constructing an estimator that satisfies these conditions.

We begin by providing assumptions upon which our first main theoretical result relies.
We consider two types of assumptions.
The first type (A) is a set of determinsitic conditions on the goodness-of-fit functional and the underlying probabilty distribution, whereas the second set of assumptions (B) is stochastic in nature and describes conditions that our estimator $\{\tilde{G}_{n,f}(\beta): f \in \mathcal{F}, \beta \in \mathcal{B}\}$ must satisfy.
\begin{itemize}
\item[] \textbf{Assumption A1: } For any $f \in \mathcal{F}$ and any $\beta \in \mathcal{B}$, $G_{P,f}(\beta)$ is pathwise differentiable in a nonparametric model, and its nonparametric efficient influence function is given by $\phi_{P,f}(\cdot; \beta): \mathcal{Z} \to \mathbb{R}$.
\item[] \textbf{Assumption A2: } $G_{P,f}$ and $\phi_{P,f}$ are twice differentiable in $\beta$ for each $f$ in $\mathcal{F}$, and the derivatives are given by 
\begin{align*}
&G'_{P,f}(\beta) := \frac{d}{d\beta} G_{P,f}(\beta), \quad
G''_{P,f}(\beta) := \frac{d^2}{d\beta^2} G_{P,f}(\beta),
\\
&\phi'_{P,f}(\cdot;\beta) := \frac{d}{d\beta} \phi_{P,f}(\cdot;\beta), \quad
\phi''_{P,f}(\beta) := \frac{d^2}{d\beta^2} \phi_{P,f}(\cdot; \beta).
\end{align*}
\item[] \textbf{Assumption A3: } There exist positive constants $C_1, C_2  > 0$ such that, for any $\{\beta_{f}: \in \mathcal{F}\}$, $\sup_{f \in \mathcal{F}} G_{0,f}(\beta_f) - G_{0,f}(\beta_{0,f}) < C_1$ implies that
\begin{align*}
\sup_{f \in \mathcal{F}}(\beta_{0,f} - \beta_f)^2 \leq C_2 \left\{\sup_{f \in \mathcal{F}} G_{0,f}(\beta_f) - G_{0,f}(\beta_{0,f})\right\}.
\end{align*}
\item[] \textbf{Assumption A4: } For each $f \in \mathcal{F}$, $G'_{0,f}(\beta_{0,f}) = 0$. Additionally, $ G''_{0,f}$ is bounded above zero in a neighborhood of $\beta_{0,f}$, uniformly in $\mathcal{F}$. That is, $\inf_{f \in \mathcal{F} } G''_{0,f}(\beta_{f})$ is positive whenever $\sup_{f \in \mathcal{F}}|\beta_f - \beta_{0,f}|$ is small.
\item[] \textbf{Assumption A5: } \sloppy Both the function classes $\left\{ \phi_{P_0,f}(\cdot;\beta): f \in \mathcal{F}, \beta \in \mathcal{B} \right\}$ and $\left\{ \phi'_{P_0,f}(\cdot;\beta_{0,f}): f \in \mathcal{F} \right\}$ are $P_0$-Donsker.
\end{itemize}

Assumption A1 requires that the goodness-of-fit is pathwise differentiable, which as noted in Section \ref{sec:plug-in}, enables us to construct $n^{1/2}$-consistent estimators.
When $G_{P,f}$ is pathwise differentiable estimand, its efficient influence function is guaranteed to exist, and knowledge of the efficient influence function is often needed for constructing efficient estimators and studying their asymptotic properties in nonparametric models.
We note that because we assume $G_{P,f_1}(0) = G_{P,f_2}(0)$ for any $f_1, f_2$ (recall we assume \eqref{submod-prop-1} holds), the efficient influence functions $\phi_{P,f_1}(\cdot;0)$ and $\phi_{P, f_2}(\cdot; 0)$ are also equal.
Assumptions A2 and A3 state that the goodness-of-fit must be smooth and convex along each of the parametric sub-models.
Assumption A4 requires that $\mathcal{B}$ is large enough to contain the global optimizer of $G_{0,f}$ over $\mathbb{R}$, and the goodness-of-fit satisfies some additional smoothness constraints in a neighborhood of the optimizer. 
Assumption A5 states that, while $\mathcal{F}$ may be specified as a large nonparametric function class, it must satisfy some mild complexity constraints.

\begin{itemize}
\item[] \textbf{Assumption B1:} For any $f_1, f_2 \in \mathcal{F}$ with $f_1 \neq f_2$ we have that $\tilde{G}_{n,f_1}(0) = \tilde{G}_{n,f_2}(0)$.
\item[] \textbf{Assumption B2:} $\tilde{G}_{n,f}$ is an asymptotically linear estimator for $G_{0,f}$ in the sense that the remainder
$\tilde{r}_{n,f}(\beta) := \left\{\tilde{G}_{n,f}(\beta) - G_{0,f}(\beta) \right\} - \frac{1}{n}\sum_{i=1}^n \phi_{0,f}(Z_i; \beta)$
satisfies 
\begin{align*}
\sup_{f\in \mathcal{F}, \beta \in \mathcal{B}} |\tilde{r}_{n,f}(\beta)| = o_P(n^{-1/2}).
\end{align*}
\item[] \textbf{Assumption B3:} The derivative of $\tilde{G}_{n,f}$ exists and is given by $\tilde{G}'_{n,f}(\beta) = \frac{d}{d\beta} \tilde{G}_{n,f}(\beta)$. Moreover, letting $\tilde{r}'_{n,f}(\beta) = \frac{d}{d\beta} r_{n,f}(\beta)$, we have
\begin{align*}
\sup_{f\in \mathcal{F}, \beta \in \mathcal{B}} |\tilde{r}'_{n,f}(\beta)| = o_P(n^{-1/2}).
\end{align*}
\item[] \textbf{Assumption B4:} The second derivative of $\tilde{G}_{n,f}$ exists and is given by $\tilde{G}''_{n,f}(\beta) = \frac{d^2}{d\beta^2} \tilde{G}_{n,f}(\beta)$. Moreover, letting $\tilde{r}''_{n,f}(\beta) = \frac{d^2}{d\beta^2} r_{n,f}(\beta)$, we have
\begin{align*}
\sup_{f\in \mathcal{F}, \beta \in \mathcal{B}} |\tilde{r}''_{n,f}(\beta)| = o_P(1).
\end{align*}
\end{itemize}

Assumption B1 states that the estimator for the goodness-of-fit along any parametric sub-model takes the same value at $\beta = 0$.
In view of condition \eqref{submod-prop-1}, all sub-models intersect and attain the same value for the goodness-of-fit at $\beta = 0$, so it is natural to assume that our estimator also has this property.
Assumptions B2 places a requirement that $\tilde{G}_{n,f}(\beta)$ is an asymptotically linear estimator for $G_{0,f}(\beta)$, where the asymptotic linearity holds uniformly over $\mathcal{F} \times \mathcal{B}$.
Assumption B3 states that $\tilde{G}_{n,f}(\beta)$ is differentiable, and the derivative $\tilde{G}_{n,f}'(\beta)$ is an asymptotically linear estimator for $G'_{0,f}(\beta)$, uniformly over $\mathcal{F} \times \mathcal{B}$.
Finally, Assumption B4 requires that the second derivative of $\tilde{G}_{n,f}(\beta)$ exists and converges in probability to the second derivative of $G''_{0,f}(\beta)$, uniformly over $\mathcal{F} \times \mathcal{B}$.

For a given estimator $\{\tilde{G}_{n,f}(\beta): f\in \mathcal{F}, \beta \in \mathcal{B}\}$ of $\{G_{0,f}: f \in \mathcal{F}, \beta \in \mathcal{B}\}$, let $\{\tilde{\beta}_{n,f}: f \in \mathcal{F}\}$ satisfy
\begin{align*}
\sup_{f\in \mathcal{F}}|\tilde{G}_{n,f}'(\tilde{\beta}_{n,f})| = o_P(n^{-1/2}).
\end{align*}
The following theorem states that, under mild regularity conditions, $\tilde{\beta}_{n,f}$ is an asymptotically linear estimator for $\beta_{0,f}$, and moreover the collection $\{\tilde{\beta}_{n,f}:f \in \mathcal{F}\}$, when appropriately standardized,  achieves a Gaussian limiting distribution.
\begin{theorem}
Let $\ell^\infty(\mathcal{F})$ denote the space of bounded functionals on $\mathcal{F}$, and let $\mathbb{H}_0$ be a tight mean zero Gaussian process with covariance
\begin{align*}
\Sigma_0: (f_1, f_2) \mapsto E_0[\phi'_{0,f_1}(Z;\beta_{0,f_1})\phi'_{0,f_2}(Z;\beta_{0,f_2})].
\end{align*}
If Assumptions A1-A4 hold, and if $\{\tilde{G}_{n,f}(\beta): f \in \mathcal{F}, \beta \in \mathcal{B}\}$ satisfies Assumptions B1-B4, then $\tilde{\beta}_{n,f}$ is asymptotically linear with influence function  
\begin{align*}
z \mapsto -\left\{ G''_{0,f}(\beta_{0,f} )\right\}^{-1} \phi'_{0,f} (z; \beta_{0,f}).
\end{align*}
\sloppy Moreover, If A5 also holds, then the process $\left\{n^{1/2}[\tilde{\beta}_{n,f} - \beta_{0,f}]: f\in \mathcal{F}\right\}$, converges weakly to $\left\{ \left[G''_{0,f}(\beta_{0,f})\right]^{-1}\mathbb{H}_0(f): f\in \mathcal{F}\right\}$ as an element of $\ell^{\infty}(\mathcal{F})$, with respect to the supremum norm.
\end{theorem}

Theorem 1 can be viewed as a generalization of well-known results that show M-estimators are asymptotically linear in finite-dimensional models \citep[see, e.g., Theorem 5.23 of ][]{van2000asymptotic}.
Our result on uniform asymptotic linearity in infinite-dimensional models can be proven using a fairly standard argument.


In what follows, we suggest some approaches for constructing an estimator that satisfies Assumptions B1-B4.
We describe at a high-level what types of conditions are needed for a given estimation strategy to be valid, though the specific requirements depend on the target estimand $G_{0,f}$ and vary from problem to problem.
Later on in Section \ref{sec:illustration1}, we demonstrate how to construct an estimator and that satisfies Assumptions B1-B4 in an example.

Suppose that one has available an estimator $\hat{P}_n$ for the underlying probability distribution $P_0$.
Typically estimation of the entire probability distribution $P_0$ it not necessary, and one will only need to estimate nuisance components upon which $G_{0,f}(\beta)$ and $\phi_{0,f}(\cdot; \beta)$ depend.
We assume that $\{G_{P,f}(\beta): f \in \mathcal{F}, \beta \in \mathcal{B}\}$ and $\{\phi_{P,f}(\cdot; \beta): f \in \mathcal{F}, \beta \in \mathcal{B} \}$ depend on $P$ only through a nuisance $Q_P$, which resides in a space $\mathcal{Q}$ endowed with norm $\| \cdot \|_{
\mathcal{Q}}$.
The true value of the nuisance component is given by $Q_{P_0}$, and the plug-in estimator for the nuisance is $Q_{\hat{P}_n}$.

As a starting point, one might consider using $G_{\hat{P}_n, f}(\beta)$ as an estimator for $G_{0,f}(\beta)$.
If $\hat{P}_n$ belongs to the model $\mathcal{M}$, the plug-in estimator satisfies Assumption B1.
This leaves Assumptions B2 through B4 to be verified.
Suppose now that $Q_P$ is itself pathwise differentiable and can therefore be estimated at an $n^{1/2}$-rate. 
Then if $Q_{\hat{P}_n}$ is an asymptotically linear estimator for $Q_{P_0}$, one can argue that $G_{\hat{P}_n,f}(\beta)$ is also be asymptotically linear by applying the delta method.
Assumption B2 then holds, as long as the asymptotic linearity is preserved uniformly over $\mathcal{F} \times \mathcal{B}$.
Asymptotic linearity of $G'_{\hat{P}_n,f}$ (Assumption B3) and consistency of $G''_{\hat{P}_n,f}$ (Assumption B4) can be established using a similar argument.

In many instances, the nuisance $Q_P$  can include quantities such as density functions or conditional mean functions which are non-pathwise differentiable in a nonparametric model.
In this case, it is not possible to construct an $n^{1/2}$-rate consistent estimator for the nuisance.
Obtaining an estimator for the nuisance usually involves making a bias variance trade-off that may be sub-optimal for the objective of estimating the goodness-of-fit.
When the nuisance estimator retains non-negligible bias, it is possible that the bias propagates, leading to $G_{\hat{P}_n,f}(\beta)$ being biased as well.
As a consequence, $G_{\hat{P}_n,f}(\beta)$ may not be asymptotically linear, and we may require more sophisticated methods to construct an $n^{1/2}$-consistent estimator.

One widely-used method for obtaining an asymptotically linear estimator when the initial estimator $G_{\hat{P}_n,f}(\beta)$ is biased is to perform a one-step bias correction \citep{pfanzagl1982contributions}.
Consider the plug-in estimator for the efficient influence function $\phi_{\hat{P},f}(\cdot; \beta)$.
The empirical average of the estimator for the efficient influence function serves as a first-order correction for the bias of the initial estimator.
By adding this empirical average to the initial estimator, one can obtain the so-called one-step estimator:
\begin{align*}
\tilde{G}_{n,f}(\beta) = G_{\hat{P}_n, f}(\beta) + \frac{1}{n}\sum_{i=1}^n \phi_{\hat{P}_n, f}(Z_i; \beta).
\end{align*}
It can be easily seen that the one-step estimator satisfies Assumption B1.
In what follows, we briefly discuss what arguments one will typically use to verify Assumptions B2-B4.
While we do not provide a detailed discussion here, we refer readers to a recent review by \cite{hines2022demystifying}, which provides a more in-depth explanation.

The estimation error of the one-step estimator has the exact representation,
\begin{align*}
\tilde{G}_{n,f}(\beta) - G_{0,f}(\beta) = \frac{1}{n} \sum_{i=1}^n \phi_{0, f}(Z_i; \beta) + R^{\mathrm{i}}_{n,f}(\beta) + R^{\mathrm{ii}}_{n,f}(\beta),
\end{align*}
where we define $R^{\mathrm{i}}_{n,f}(\beta)$ and $R^{\mathrm{ii}}_{n,f}(\beta)$ as
\begin{align*}
&R^{\mathrm{i}}_{n,f}(\beta) := \frac{1}{n}\sum_{i=1}^n \left\{\phi_{\hat{P}_n, f}(Z_i; \beta) - 
\phi_{0, f}(Z_i; \beta) \right\} -  \int \left\{ \phi_{\hat{P}_n, f}(z; \beta) - \phi_{0, f}(z; \beta) \right\} dP_0(z),
\\
&R^{\mathrm{ii}}_{n,f}(\beta) := \left\{ G_{\hat{P}_n, f}(\beta) - G_{0,f}(\beta) \right\} + \int \phi_{\hat{P}_n, f}(z; \beta) dP_0(z).
\end{align*}
Asymptotic linearity of the one-step estimator follows if it can be established that $R^{\mathrm{i}}_{n,f}(\beta)$ and $R^{\mathrm{ii}}_{n,f}(\beta)$ converge to zero in probability at an $n^{1/2}$-rate.
The first term $R^{\mathrm{i}}_{n,f}(\beta)$ is a difference-in-differences remainder that is asymptotically negligible when $\phi_{\hat{P}_n,f}(\cdot; \beta)$ is consistent for $\phi_{P_0,f}(\cdot; \beta)$ and $\phi_{\hat{P}_n,f}(\cdot; \beta)$ is contained within a $P_0$-Donsker class \citep[see Lemmas 19.24 and 19.26 of][]{van2000asymptotic}.
The second term $R^{\mathrm{ii}}_{n,f}(\beta)$ is a second-order remainder term, which can usually be bounded above by the squared norm of the difference between the nuisance estimator and its true value, $\| Q_{\hat{P}_n} - Q_{P_0} \|_{\mathcal{Q}}^2$.
One can argue that if the nuisance estimator is $n^{1/4}$-consistent with respect to $\| \cdot \|_{\mathcal{Q}}$, then $R^{\mathrm{ii}}_{n,f}(\beta) = o_P(n^{-1/2})$.
Even in a nonparametric model, there exist several approaches for constructing $n^{1/4}$-rate consistent nuisance estimators when one makes only mild structural assumptions on $Q_{P_0}$, such as smoothness or monotonicity \citep[see, e.g.,][]{geer2000empirical, tsybakov2009nonparametric}.
To verify that Assumptions B3 and B4 hold, one can perform a similar analysis to show that the first and second derivatives of the remainder terms $R^{\mathrm{i}}_{n,f}(\beta)$ and $R^{\mathrm{ii}}_{n,f}(\beta)$, with respect to $\beta$, tend to zero at the requisite rate.

While we focused on one-step estimation above because we find its simplicity appealing, other strategies for constructing of bias-corrected estimators, such as targeted minimum-loss based estimation could alternatively be used.
These strategies are usually also viable under a similar set of regularity conditions.

\subsection{Asymptotic properties of proposed estimator} \label{sec:asymp}

We are at this point prepared to describe the bias-corrected estimator for $\Psi_0$ and its asymptotic properties.
As stated in Section \ref{sec:plug-in}, we estimate $\Psi_0$ as  
\begin{align}
\tilde{\Psi}_n = \sup_{f \in \mathcal{F}} \tilde{G}_{n,f}(0) - \tilde{G}_{n,f}(\tilde{\beta}_{n,f}),
\label{iif-estimator}
\end{align}
where $\{\tilde{G}_{n,f}(\beta):f \in \mathcal{F}, \beta \in \mathcal{B}\}$ is an estimator satisfying the conditions outlined in Section \ref{sec:submodel-inference}.

In this section we establish weak convergence of $\tilde{\Psi}_n$.
We show that $\tilde{\Psi}_n$ attains a tractable limiting distribution under mild regularity conditions, but the limiting distribution and convergence rate depend on the true value of $\Psi_0$.
We study two cases.  
First, we consider the setting in which $\Psi_0 = 0$, and the null hypothesis of no improvement in fit \eqref{null} holds.
Second, we study the case in which $\Psi_0$ is a positive constant.

\noindent \textbf{Case 1: The improvement in fit is zero ($\Psi_0 = 0$)}

Suppose that the null of no improvement in fit holds.
Recall from Section \ref{sec:prelim} that when $\Psi_0 = 0$, $\sup_{f \in \mathcal{F}}|\beta_{0,f}| = 0$.
Also, as discussed in Section \ref{sec:plug-in}, by performing a Taylor expansion for $\tilde{G}_{n,f}$ around $\tilde{\beta}_{n,f}$, we have 
\begin{align}
\tilde{\Psi}_{n} = \frac{1}{2}\sup_{f \in \mathcal{F}} G''_{n,f}(\check{\beta}_{n,f}) \tilde{\beta}^2_{n,f},
\label{Case1-expansion}
\end{align}
for some $\check{\beta}_{n,f}$ satisfying $|\check{\beta}_{n,f} - \beta_{0,f}| \leq |\tilde{\beta}_{n,f} - \beta_{0,f}|$.
Under Assumption B4, we are able, in \eqref{Case1-expansion}, to replace $G''_{n,f}(\check{\beta}_{n,f})$ with $G''_{0,f}(\beta_{0,f})$.
This and the fact that $\sup_{f\in\mathcal{F}} \beta_{n,f}^2 = O_P(n^{-1})$ allow us to write
\begin{align}
\tilde{\Psi}_{n} &= \frac{1}{2}\sup_{f \in \mathcal{F}} G''_{0,f}(0) \tilde{\beta}^2_{n,f} + o_P(n^{-1})
= \sup_{f \in \mathcal{F}}\frac{1}{2 G''_{0,f}(0)} \left[ \frac{1}{n}\sum_{i=1}^n \phi'_{P_0, f}(Z_i; 0)\right]^2 + o_P(n^{-1}). \label{null-representation}
\end{align}
Thus, under the null, $\tilde{\Psi}_n$ can be represented as the squared supremum of an empirical process, plus an asymptotically negligible remainder.  
By applying Theorem 1 in conjunction with Slutsky's theorem and the continuous mapping theorem, we have that $n \tilde{\Psi}_{n}$ converges weakly to \sloppy $\sup_{f \in \mathcal{F}} \left[\left\{2 G''_{0,f}(0)\right\}^{-1/2}\mathbb{H}(f)\right]^2$, where $\mathbb{H}$ is the Gaussian process described in Theorem 1.
The following Theorem states this result formally.

\begin{theorem}
Suppose that the null hypothesis of no improvement in fit \eqref{null} holds, and the Assumptions of Theorem 1 are all satisfied.
Then $\tilde{\Psi}_n$ converges weakly to $\sup_{f \in \mathcal{F}} \left[\left\{2 G''_{0,f}(0)\right\}^{-1/2}\mathbb{H}(f)\right]^2$.
\label{thm-Case1Limit}
\end{theorem}

We can apply Theorem \ref{thm-Case1Limit} to obtain an approximation for the sampling distribution of $\tilde{\Psi}_n$ under the null of zero improvement in fit,  making it easy for us to perform a hypothesis test.
In particular, Theorem \ref{thm-Case1Limit} implies that a test which rejects the null when $n\tilde{\Psi}_n$ is larger than the $(1 - \alpha)$ quantile of the distribution of $\sup_{f \in \mathcal{F}} \left[\left\{2 G''_{0,f}(0)\right\}^{-1/2}\mathbb{H}(f)\right]^2$ will achieve type-1 error control at the $\alpha$-level in the limit of large $n$.

We used two key ingredients to construct an improvement in fit estimator with parametric rate convergence under the null.  
First, we found it useful to the represent the difference between the full and reduced models for the function-valued parameter of interest as the union of many one-dimensional parametric sub-models.
We have deduced that, under the null, the asymptotic behavior of an improvement in fit estimator is determined in large part by the complexity of the collection of paths along which the estimated goodness-of-fit minimizer over the full model can possibly approach the minimizer over the reduced model.
We found it necessary to constrain the complexity to ensure that the improvement in fit estimator converges sufficiently quickly.
In our regime, this can be easily achieved by restricting the size of $\mathcal{F}$.
We expect that if one were to assume a different form for $\Theta \setminus \Theta^*$, one would still need to impose a constraint that plays a similar role in order to obtain an $n$-rate consistent estimator.
Second, efficient estimation of the improvement in fit along any sub-model is needed.
In settings where the reduced model is infinite-dimensional, estimation of the goodness-of-fit minimizer over the reduced model can generate bias for the improvement in fit estimator and reduce its convergence rate.
Fortunately, an efficient estimator can be obtained using standard techniques for bias correction.


\noindent \textbf{Case 2: The improvement in fit is bounded away from zero $(\Psi_0 > 0)$}

Now, consider the setting where $\Psi_ 0$ is a positive constant.
Let $f_0$ and $f_n$ be functions that satisfy
\begin{align}
G_{0, f_0}(\beta_{0,f_0}) = \sup_{f \in \mathcal{F}} G_{0,f}(\beta_{0,f}), \quad
G_{n, f_n}(\tilde{\beta}_{n,f_n}) = \sup_{f \in \mathcal{F}} G_{0,f}(\tilde{\beta}_{n,f}).
\label{least-fav-path}
\end{align}
We can express the estimation error of $\tilde{\Psi}_n$ as
\begin{align*}
\tilde{\Psi}_n - \Psi_0 &= 
\left\{\sup_{f_1 \in \mathcal{F}}G_{n,f_1}(0) -  G_{n,f_1}(\beta_{n,f_1})\right\} -  \left\{\sup_{f_2 \in \mathcal{F}} G_{0,f_2}(0) - G_{0,f_2}(\beta_{0,f_2})\right\}
\\
&= \left\{ \tilde{G}_{n,f_n}(0) - \tilde{G}_{n,f_n}(\beta_{n,f_n}) \right\} -  \left\{G_{0,f_0}(0) - G_{0,f_0}(\beta_{0,f_0}) \right\}.
\end{align*}
One might expect that $f_n$ should approach $f_0$ as $n$ grows, so $G_{n,f_n}(\tilde{\beta}_{n,f_n})$ should behave similarly to $G_{n,f_0}(\beta_{0,f_0})$.
In fact, if one could establish that
\begin{align}
\left|\left\{\tilde{G}_{n,f_0}(0) - \tilde{G}_{n,f_0}(\beta_{0,f_0})\right\} -  \left\{\tilde{G}_{n,f_n}(0) -  \tilde{G}_{n,f_n}(\tilde{\beta}_{n,f_n})\right\}\right| = o_P(n^{-1/2}),
\label{case-2-remainder}
\end{align}
then they would be able to conclude that $\tilde{\Psi}_n$ is asymptotically linear with influence function $z \mapsto \{\phi_{P_0, f_0}(z; 0) - \phi_{P_0, f_0}(z; \beta_{0,f_0})\}$ under Assumption B2.

The remainder term in \eqref{case-2-remainder} is $o_P(n^{-1/2})$ under mild assumptions.
Because $\tilde{G}_{n,f_0}(0) - \tilde{G}_{n,f_n}(0)$ is zero under Assumption B1,  it only needs to be shown that $\tilde{G}_{n,f_0}(\beta_{0,f_0}) - \tilde{G}_{n,f_n}(\tilde{\beta}_{n,f_n})$ is asymptotically negligible. 
Because the goodness-of-fit estimator is asymptotically linear, $\tilde{G}_{n,f_0}(\beta_{0,f_0}) - \tilde{G}_{n,f_n}(\tilde{\beta}_{n,f_n})$ is approximately equal to $G_{0,f_0}(\beta_{0,f_0}) - G_{0,f_n}(\beta_{0, f_n})$, which is commonly referred to as the \textit{excess risk} in the literature on M-estimation \citep{geer2000empirical}.
Thus, in essence, one can verify \eqref{case-2-remainder} by showing that the excess risk converges to zero in probability at an $n^{1/2}$-rate.
This can be done using standard arguments from the M-estimation literature.
The following result provides explicit conditions under which $\tilde{\Psi}_n$ is an asymptotically linear estimator for $\Psi_0$.
\begin{theorem}
Suppose that the improvement in fit is positive, i.e., $\Psi_0 > 0$.
Suppose further that Assumptions A1, A5, B1, and B2 hold, and there exists a sequence $d_n = o(n^{1/2-\delta})$ for some $\delta > 0$ such that
\begin{align}
\sup_{\{(f,\beta): G_{0,f}(\beta) - G_{0,f_0}(\beta_{0,f_0}) \leq d_n \}}\left[\int \left\{\phi_{P_0,f}(z;\beta) - \phi_{P_0,f_0}(z; \beta_{0,f}) \right\}^2dP_0(z)\right]^{1/2} = o(1).
\label{margin}
\end{align}
Then $\tilde{\Psi}_n$ is an asymptotically linear estimator for $\Psi_0$ with influence function 
\begin{align*}
z \mapsto \phi_{P_0, f_0}(z; 0) - \phi_{P_0,f_0}(z;\beta_{0,f_0}).
\end{align*}
\end{theorem}

An important consequence of Theorem 3 is that $\tilde{\Psi}_n$ is asymptotically efficient in a nonparametric model, and hence performs as well as the plug-in estimator described in Section 3 when $\Psi_0 > 0$.
The assumption in \eqref{margin} is a type of smoothness condition that is assumed commonly in the literature on estimation in high-dimensional and nonparametric models 
\citep[see, e.g.,][]{van2008high, negahban2012unified, bibaut2019fast}.
The condition ensures that $\phi_{P_0,f}(z;\beta)$ and $\phi_{P_0,f_0}(z;\beta_{0,f_0})$ are close in $L_2(P_0)$ distance when $G_{0,f}(\beta) - G_{0,f}(\beta_{0,f})$ is small.

Some conditions that are needed by Theorem 2 are not needed by Theorem 3.
Notably, it is not necessary for $\sup_{f \in \mathcal{F}} |G'_{0,f}(\beta_{0,f})|$ to be zero.
This means that $\mathcal{B}$ can be mis-specified in the sense that the interval is too small, and along any sub-model $\theta^*_{P,f}$, there can exist a candidate that achieves a better fit than $\theta^*_{P,f}(\cdot;\beta_{0,f})$.
In other words, we allow there to be $\beta_{0,f}^* \in \mathbb{R}\setminus \mathcal{B}$ for which $G_{0,f}(\beta_{0,f}) > G_{0,f}(\beta^*_{0,f})$.  
Even then, $\tilde{\Psi}_n$ remains an asymptotically linear estimator for $\sup_{f \in \mathcal{F}, \beta \in \mathcal{B}} \{G_{0,f}(0) - G_{0,f}(\beta)\}$.

\section{Construction of tests and intervals for the improvement in fit} \label{sec:bootstrap}

In this section, we propose strategies for testing and confidence set construction for the improvement in fit.
Our approach uses a computationally efficient bootstrap algorithm, which we describe in detail below.

We also provide theoretical results that establish validity of our proposed bootstrap method.
Before proceeding, it is helpful to first state regularity conditions upon which our result rely.

\begin{itemize}
\item[] \textbf{Assumption C1: } The function class $\left\{ \left[G''_{P,f}(0) \right]^{-1/2} \phi'_{P,f}(\cdot;0): f \in \mathcal{F}\right\}$ depends on $P$ only through a nuisance $Q$, which takes values in a space $\mathcal{Q}$ endowed with norm $\|\cdot\|_{\mathcal{Q}}$, and our estimator $Q_{\hat{P}_n}$ satisfies $\| Q_{\hat{P}_n}  - Q_{P_0}\|_{\mathcal{Q}} = o_P(1)$.
\item[] \textbf{Assumption C2: }   $\|Q_P - Q_{P_0} \|_{\mathcal{Q}}$ approaches zero, both $\sup_{f \in \mathcal{F}} \int \left\{\phi_{P,f}'(z; 0) - \phi'_{P_0, f}(z; 0) \right\}^2dP_0(z)$ and $\sup_{f \in \mathcal{F}}|G''_{P,f}(0) - G_{P_0,f}''(0)|$ tend to zero as well.
\item[] \textbf{Assumption C3: } There exist $\delta_1, \delta_2 > 0$ such that the function classes
\begin{align*}
&\Phi_{\delta_1} := \left\{\phi_{P,f}(\cdot;0) - \phi_{P,f}(\cdot;\beta): f \in \mathcal{F}, \beta \in \mathcal{B}, \|Q_P - Q_{P_0}\| < \delta_1 \right\} ,
\\
&\Phi'_{\delta_2} := \left\{\phi'_{P,f}(\cdot;0) : f \in \mathcal{F}, \|Q_P - Q_{P_0}\| < \delta_2 \right\},
\end{align*}
are $P_0$-Donsker, with finite squared envelope function and finite bracketing integral \citep[see, e.g., Chapter 19 of][]{van2000asymptotic}, and 
\begin{align*}
\inf\left\{ G''_{P,f}(0): f \in \mathcal{F}, \|Q_P - Q_{P_0} \| < \delta_2\right\} > 0.
\end{align*}
\end{itemize}

Assumption C1 states that, for our bootstrap methods to be viable, estimation of the entire probability distribution is not needed, and it is sufficient to only estimate nuisance parameters upon which the efficient influence function depends.  
Recall that we made a similar assumption when we described construction of asymptotically linear estimators in Section \ref{sec:submodel-inference}.
Assumption C2 states that when we estimate the nuisance components consistently, the plug-in estimator for the efficient influence function is consistent as well.
Assumption C3 states that our efficient influence function estimator belongs to a function class that is not overly complex, with probability tending to one.

\subsection{Approximation of the null limiting distribution} \label{sec:bs-test}

To perform a test of the hypothesis of no improvement in fit, we need an approximation for the asymptotic cumulative distribution function of $\tilde{\Psi}_n$ under the null.
While we are able to characterize the null limiting distribution of $\tilde{\Psi}_n$ using Theorem 2, it is possible that a closed form expression for the distribution function is not available.
However, we can use resampling techniques to obtain an approximation.

We approximate the null limiting distribution of $\tilde{\Psi}_n$ using the multiplier bootstrap method proposed by \cite{hudson2021inference}.
The multiplier bootstrap is a computationally efficient method for approximating the sampling distribution estimators that can be represented as a functional of a well-behaved empirical process, plus a negligible remainder.
Such an approach is applicable in our setting because $\tilde{\Psi}_n$ has such representation (see \eqref{null-representation}).

For $m = 1, 2, \ldots, M$ and $M$ large, let $\boldsymbol{\xi}_m = (\xi_{1,m}, \ldots, \xi_{n,m})$ be an $n$-dimensional vector of independent Rademacher random variables, also drawn independently from $Z_1, \ldots, Z_n$.
We define the multiplier bootstrap statistic
\begin{align}
T_{n,m}^\xi := \sup_{f \in \mathcal{F}}\frac{1}{2G''_{\hat{P}_n,f}(0)}\left\{\frac{1}{n}\sum_{i=1}^n \xi_{i,m}\phi_{\hat{P}_n, f}(Z_i; 0)\right\}^2,
\label{mult-bs}
\end{align}
as an approximate draw from the null limiting distribution of $\tilde{\Psi}_n$.

For a realization $t$ of $n\tilde{\Psi}_n$, let 
\begin{align}
\rho_{0}(t) :=  P_0\left(\sup_{f \in \mathcal{F}} \frac{1}{2G''_{0,f}(0)}\mathbb{H}^2(f) > t\right) = \lim_{n \to \infty}P_0(\tilde{\Psi}_n > n^{-1}t),
\label{p-value}
\end{align}
denote the p-value for a test of no improvement in fit, based on the limiting distribution of $\tilde{\Psi}_n$.
Given a large sample of multiplier bootstrap statistics, one can approximate the p-value as
\begin{align*}
\rho_{M,n}(t) := \frac{1}{M} \sum_{m = 1}^M \mathds{1}\left(T^{\xi}_{n,m} > n^{-1}t \right).
\end{align*}
The following result due to \cite{hudson2021inference} provides conditions under which the bootstrap approximation of the limiting distribution is asymptotically valid, and use of the bootstrap p-value is appropriate.

\begin{theorem}
Let $\xi_1,\xi_2,\ldots,\xi_n$ be independent Rademacher random variables, also independent of $Z_1, \ldots, Z_n$, and let $T_{n}^\xi = \sup_{f \in \mathcal{F}}\frac{1}{2G''_{\hat{P}_n,f}(0)}\left\{\frac{1}{n}\sum_{i=1}^n \xi_{i}\phi_{\hat{P}_n, f}(Z_i; 0)\right\}^2$.
Under Assumptions C1 through C3, $nT_n^\xi$ converges weakly to  converges weakly to $\sup_{f \in \mathcal{F}} \left[\left\{2 G''_{0,f}(0)\right\}^{-1/2}\mathbb{H}(f)\right]^2$, conditional upon $Z_1, \ldots, Z_n$, in outer probability.  
\end{theorem}

\subsection{Interval construction for $\Psi_0$} \label{sec:bs-interval}

In this section, we present a method for constructing a confidence interval for $\Psi_0$.
The standard approach for interval construction based on a Gaussian approximation of the sampling distribution of an estimator is inadvisable because $\tilde{\Psi}_n$ is only asymptotically Gaussian when $\Psi_0$ is bounded away from zero.
We show that this issue can be overcome by instead constructing a confidence interval via hypothesis test inversion.

Suppose that one could perform a level $(1-\alpha)$ level test of the hypothesis $H: \Psi_0 = \psi$ for any $\psi \geq 0$.
Then the set
\begin{align*}
\mathcal{C}_n^{1-\alpha} := \left\{\psi \geq 0: \text{We fail to reject } \Psi_0 = \psi \text{ based on } Z_1, \ldots, Z_n \right\},
\end{align*}
would be a $100(1-\alpha)\%$ confidence interval for $\Psi_0$. That is, in the limit of large $n$, $\mathcal{C}_n^{1-\alpha}$ would contain $\Psi_0$ with probability at least $(1-
\alpha)$.

We construct a test of $\Psi_0 = \psi$ using the test statistic $S_n(\psi) := |\tilde{\Psi}_n - \psi|$. 
Let $s^{1-\alpha}_n$ be an approximation for the $(1-\alpha)$ quantile of the limiting distribution of $|\tilde{\Psi}_n - \Psi_0|$.
For a suitable $s^{1-\alpha}_n$, a test that rejects the null when $S_n(\psi)$ exceeds $s^{1-
\alpha}_n$ will achieve asymptotic type-1 error control at the level $(1-\alpha)$.
Moreover, an asymptotically valid confidence set can be obtained by setting
\begin{align*}
\mathcal{C}_n^{1-\alpha} &= \left\{\psi \geq 0: S_n(\psi) \leq s^{1-\alpha}_n \right\}
= 
\left[\max\left(0, \tilde{\Psi}_n - s_n^{1-\alpha}\right), \tilde{\Psi}_n + s_n^{1 - \alpha} \right].
\end{align*} 

It is not immediately obvious how to select $s_n^{1-\alpha}$ because the limiting distribution and convergence rate of $|\tilde{\Psi}_n - \Psi_0|$ depend on whether $\Psi_0 = 0$.
To address this concern, in what follows, we present a multiplier bootstrap approximation of the limiting distribution that adapts to the unknown value of $\Psi_0$.

Let $\pi_n$ be any random sequence that converges to one in probability when $\Psi_0 = 0$, and converges to zero in probability to when $\Psi_0 > 0$.  
For instance, we can set
\begin{align}
\pi_n = \rho_{M,n}\left( \frac{n}{\log(n) } \tilde{\Psi}_n \right), 
\label{pi}
\end{align}
where $\rho_{M,n}$ is the multiplier bootstrap p-value.
That this choice of $\pi_n$ is valid follows from the fact that $\tilde{\Psi}_n$ is consistent for $\Psi_0$ and $n$-rate convergent under the null.
%
Now, similar to Section 5.1, for $m = 1, \ldots, M$ and $M$ large, we generate a pair of random variables as follows.
The first random variable is $T_{n,m}^\xi$ in \eqref{mult-bs}, which is a multiplier bootstrap approximation of a draw from the limiting distribution of $|\tilde{\Psi}_n - \Psi_0|$ under the setting where $\Psi_0 = 0$.
We take the second random variable as a multiplier bootstrap approximation of a draw from the limiting distribution of $|\tilde{\Psi}_n - \Psi_0|$ when $\Psi_0 > 0$.
Specifically, we define this second random variable as
\begin{align}
U^{\xi}_{n,m} := \left|\frac{1}{n}\sum_{i=1}^n \xi_{i,m} \left\{\phi_{\hat{P}_n, f_n}(Z_i; 0) - \phi_{\hat{P}_n, f_n}(Z_i; \tilde{\beta}_{n, f_n})\right\} \right|,
\label{mult-bs-alt}
\end{align}
where $\boldsymbol{\xi}_m$ is the vector of Rademacher random variables defined in Section \ref{sec:bs-test} (the same vector may be used to construct $T_{n,m}^\xi$ and $U_{n,m}^\xi$), and $f_n$ is as defined \eqref{least-fav-path}.
Finally, we take an approximate draw from the sampling distribution of $|\tilde{\Psi}_n - \Psi_0|$ as $V_{n,m}^\xi$, where
\begin{align*}
V_{n,m}^\xi := \pi_n T^{\xi}_{n,m} + (1 - \pi_n) U^{\xi}_{n,m},
\end{align*}
and we set $s_n^{1-\alpha}$ as the $(1 - \alpha)$ quantile of $(V_{n,1}^\xi, \ldots V_{n,M}^\xi)$.

Because $\pi_n$ converges to zero when $\Psi_0$ is zero and approaches one when $\Psi_0$ is large, $V_{n,m}^\xi$ adaptively identifies whether $T^\xi_{n,m}$ or $U^\xi_{n,m}$ is a more appropriate approximation of a draw from the sampling distribution of $|\tilde{\Psi}_n - \Psi_0|$.
The following result states that $V_{n,m}^\xi$ is an asymptotically valid approximation regardless of whether $\Psi_0$ is zero or nonzero, thereby justifying our selection of $s_n^{1-\alpha}$.
\begin{theorem}
Let $\xi_1,\xi_2,\ldots,\xi_n$ be independent Rademacher random variables, also independent of $Z_1, \ldots, Z_n$.
Let $\pi_n$ be a random sequence that converges to one in probability when $\Psi_0 = 0$ and converges to zero in probability when $\Psi_0 > 0$.
Let $T_{n}^\xi = \sup_{f \in \mathcal{F}}\frac{1}{2G''_{\hat{P}_n,f}(0)}\left\{\frac{1}{n}\sum_{i=1}^n \xi_{i}\phi_{\hat{P}_n, f}(Z_i; 0)\right\}^2$, let $U_{n}^\xi =\left|\frac{1}{n}\sum_{i=1}^n \xi_{i} \left\{\phi_{\hat{P}_n, f_n}(Z_i; 0) - \phi_{\hat{P}_n, f_n}(Z_i; \tilde{\beta}_{n, f_n})\right\} \right|$, and $V_n^\xi = \pi_n T_n^\xi + (1 - \pi_n) U_n^\xi$.
Let $\mathbb{I}$ be a mean zero Gaussian random variable with variance $E_0[\{\phi_{P_0,f_0}(Z; 0) - \phi_{P_0,f_0}(Z;\beta_{0,f_0})\}^2]$, with $f_0$ defined in \eqref{least-fav-path}.
Suppose that Assumptions C1-C3 are met.
Then when $\Psi_0 = 0$, and the conditions of Theorem 2 hold, $nV_n^\xi$ converges weakly to  converges weakly to $\sup_{f \in \mathcal{F}} \left[\left\{G''_{0,f}(0)\right\}^{-1/2}\mathbb{H}(f)\right]^2$, conditional upon $Z_1, \ldots, Z_n$, in outer probability.
When $\Psi_0 > 0$, $n^{1/2}V_n^\xi$, and the conditions of Theorem 3 hold, converges weakly to $|\mathbb{I}|$, conditional upon $Z_1, \ldots, Z_n$, in outer probability.
\end{theorem}


%
%

\section{Implementation} \label{sec:implementation}

In this section we discuss implementation of our proposed method for inference on the improvement in fit.
First we describe how to construct a model for $\Theta \setminus \Theta^*$.
We subsequently discuss how to calculate the improvement in fit estimator and how to implement our proposed bootstrap procedures for testing the null of no improvement in fit and constructing confidence sets.

\subsection{Constructing the collection of parametric sub-models} \label{sec:implement-class}


We propose to construct $\mathcal{F}$ as a space of linear combinations of basis functions from $\mathcal{O}$ to $\mathbb{R}$, where the coefficients for the basis functions are required to satisfy a constraint that induces structure on the function class.
Let $\mathcal{H} = h_1 \oplus h_2 \oplus \cdots$ be a vector space defined as the span of basis functions $h_1, h_2, \ldots,$ from $\mathcal{O}$ to $\mathbb{R}$.
Let $\Gamma$ be a functional on $\mathcal{H}$ that measures the complexity of any function in $\mathcal{H}$, with larger values corresponding to greater complexity.
We set $\mathcal{F}$ to have bounded complexity.
Additionally, we impose a constraint that $\inf_{f \in \mathcal{F}} G''_{0,f}(0)$ is bounded away from zero.
In view of Assumption A4, such a constraint is needed in order for us to establish weak convergence of our proposed improvement in fit estimator under the null.
Finally, we set $\mathcal{F} = \mathcal{F}_{\lambda}$, where we define
\begin{align}
\mathcal{F}_{\lambda} := \left\{ f = \sum_{j=1}^\infty a_j h_j: a_1, a_2, \ldots \in \mathbb{R}, \frac{\Gamma(f)}{G''_{0,f}(0)} \leq \lambda \right\},
\label{implementation-class}
\end{align}
and $\lambda$ is a tuning parameter.
In practice, we recommending truncating the basis at a large level $J$ to facilitate computation.

As an example, one could construct $\mathcal{F}$ using a reproducing kernel Hilbert space (RKHS). 
Let $\kappa: \mathcal{O} \times \mathcal{O} \to \mathbb{R}$ be a positive definite kernel function, and let $\mathcal{S}_\kappa$ denote its unique reproducing kernel Hilbert space, endowed with inner product $\langle \cdot, \cdot \rangle_\kappa$.
One can select the basis functions $h_1, h_2, \ldots$ as the eigenfunctions of $\kappa$, with respect to the RKHS inner product.
We denote the corresponding eigenvalues by $\gamma_1 \leq \gamma_2 \leq \ldots$.
The complexity of any function $s:o \mapsto \sum_{j=1}^\infty a_j h_j(o)$ can be measured by its RKHS norm $\Gamma(s) = \langle s, s \rangle_\kappa = \sum_j \left( \frac{a_j}{\gamma_j} \right)^2$.  
The RKHS norm is a measure of smoothness, with higher values corresponding with lesser smoothness.
Reproducing kernel Hilbert spaces are appealing because they are flexible and contain close approximations of smooth functions \citep{micchelli2006universal}.
Moreover, that the RKHS norm is available in quadratic form in the coefficients simplifies computation.
Alternative approaches, such as constructing $\mathcal{H}$ using a spline basis and setting $\Gamma$ as a variation norm, are commonly used in nonparametric regression problems and could also be considered \citep[see, e.g.,][]{tibshirani2005sparsity, benkeser2016highly}. 

We now discuss specification of the interval $\mathcal{B}$.
The choice of $\mathcal{B}$ does not affect the null limiting distribution but may affect the limiting distribution when $\Psi_0 > 0$.
The main role of $\mathcal{B}$ is to regularize $\sup_{f \in \mathcal{F}}|\tilde{\beta}_{n,f}|$ to ensure that variance of the estimator is well-controlled.
Recall from our discussion of Theorem 3 in Section \ref{sec:asymp} that $\beta_{0,f}$, which is defined to be the minimizer of $G_{P_0,f}$ over $\mathcal{B}$, does not need to be the global minimizer over $\mathcal{B}$; our results show that $\tilde{\Psi}_n$ has a well-behaved limiting distribution regardless.
We treat the width of the interval as an additional tuning parameter.

We find that in some settings, $\tilde{\Psi}_n$ can retain good asymptotic behavior under the alternative even when $\mathcal{B}$ is taken to be an interval of arbitrary width.
In view of Theorem 1, the variance of $\tilde{\beta}_{n,f}$ has an inverse relationship with $G''_{0,f}(\beta_{0,f})$.  
Therefore, constructing $\mathcal{F}$ to only include functions for which $G''_{0,f}(\beta_{0,f})$ is bounded from below also serves to regularize $\sup_{f \in \mathcal{F}}|\tilde{\beta}_{n,f}|$. 
In particular, when $G_{0,f}$ is a quadratic function, and $G''_{0,f}$ is a constant function, it is sufficient to ensure that $G''_{0,f}(0)$ away from zero.
Because this constraint is already incorporated into $\mathcal{F}$ with the above specification, constraining the width of $\mathcal{B}$ is unnecessary in such instances.

We recommend selecting the tuning parameters  $\lambda$, and $\mathcal{B}$ (when needed) by performing cross-validation with respect to the loss $f \mapsto \tilde{G}_{n,f}(\tilde{\beta}_{f,n})$.
We note that while our asymptotic results implicitly assume that $\mathcal{F}$ is pre-specified, it is argued in \cite{hudson2021inference} that one can select $\mathcal{F}$ data-adaptively without compromising type-1 error control as long as the adaptive choice converges to a fixed class.
In some settings, e.g., when the sample size is small, it is possible that the data-adaptive choice is highly or moderately variable, and that failure to account for this variability could lead to type-1 error inflation.
One can avoid this issue by using a more conservative sample splitting approach, wherein one partition of the data is used for tuning parameter selection, and a second independent partition is used to estimate $\Psi_0$.

\subsection{Computation} \label{sec:comp}

We now discuss how to calculate the improvement in fit estimator $\tilde{\Psi}_n$ and how to implement the multiplier bootstrap for hypothesis testing and confidence interval construction.

Calculating $\tilde{\Psi}_n$ requires us to solve the optimization problem in \eqref{iif-estimator}.
When we use the specification of $\mathcal{F}$ in Section \ref{sec:implement-class}, it is possible for this problem to be non-convex, and it can be particularly challenging to solve when a closed form solution for $\tilde{\beta}_{n,f}$ is not available.
We find, however, that when $\tilde{G}_{n,f}$ is a quadratic function of $\beta$, a computationally efficient solution is available.
In Examples 1 and 3 in  Section \ref{sec:examples}, $G_{P,f}$ is a quadratic function when one considers a sub-model of the form $\theta^*_{P,f}(\cdot; \beta) = \theta^*_P(\cdot) + \beta f(\cdot)$, so this special case captures at least some examples.
In what follows, we present an approach for solving this the problem in the setting where $\tilde{G}_{n,f}$ is a quadratic function of $\beta$.
We then describe a more general method in the Supplementary Materials.

Suppose that for any $f = \sum_{j=1}^J a_j h_j$ and $\mathbf{a} = (a_1, \ldots, a_J)$, there exists a $J \times J$-dimensional matrix $\mathbf{H}_1$ and a $J$-dimensional vector $\mathbf{H}_2$ such that
\begin{align*}
\tilde{G}_{n,f}(\beta) = \beta^2 \mathbf{a}^\top \mathbf{H}_1 \mathbf{a} - 2\beta \mathbf{H}_2^\top \mathbf{a} + \text{const},
\end{align*}
where ``const'' refers to a constant that depends neither on $\mathbf{a}$ nor $\beta$.
It can be easily seen that $\tilde{\beta}_{n,f}$ has the exact representation
\begin{align*}
\tilde{\beta}_{n,f} = \frac{\mathbf{H}^\top_2\mathbf{a}}{\mathbf{a}^\top \mathbf{H}_1 \mathbf{a}}.
\end{align*}
Additionally, the second derivative estimator satisfies $\tilde{G}_{n,f}(\beta) = \mathbf{a}^\top \mathbf{H}_1 \mathbf{a}$ for all $\beta$.
Now, $\tilde{G}_{n,f}(0) - \tilde{G}_{n,f}(\tilde{\beta}_{n,f})$ can be expressed as
\begin{align*}
2\left\{\tilde{G}_{n,f}(0) - \tilde{G}_{n,f}(\tilde{\beta}_{n,f})\right\} = \frac{\mathbf{a}^\top\mathbf{H}_2^\top \mathbf{H}_2 \mathbf{a}}{\mathbf{a}^\top \mathbf{H}_1 \mathbf{a}}.
\end{align*}

Suppose now that $\Gamma(f)$ is available in quadratic form in the coefficients of the basis functions -- that is, $\Gamma(f) = \mathbf{a}^\top \mathbf{L} \mathbf{a}$ for a $J\times J$ matrix $\mathbf{L}$.
Using the above representation for $\tilde{G}_{n,f}$, we can express $\tilde{\Psi}_n$ as
\begin{align}
2\tilde{\Psi}_n &= \max_{\mathbf{a}} \left\{ \frac{\mathbf{a}^\top\mathbf{H}_2\mathbf{H}_2^\top \mathbf{a}}{\mathbf{a}^\top \mathbf{H}_1 \mathbf{a}}: \frac{\mathbf{a}^\top \mathbf{L} \mathbf{a}}{\mathbf{a}^\top \mathbf{H}_1 \mathbf{a}} \leq \lambda, \right\} \nonumber
\\
&= \max_{\mathbf{a}} \left\{ \mathbf{a}^\top\mathbf{H}_2\mathbf{H}_2^\top \mathbf{a}: \mathbf{a}^\top \mathbf{L} \mathbf{a} \leq \lambda, \mathbf{a}^\top \mathbf{H}_1 \mathbf{a} \leq 1 \right\}. \label{comp-iif}
\end{align}
The optimization problem in \eqref{comp-iif} is a quadratically constrained quadratic program (QCQP) and can be solved using publicly available software, such as the CVXR package in R \citep{fu2017cvxr}.

Multiplier bootstrap samples can be calculated using a similar method.
We first observe when we use the specification of $\theta^*_{P,f}(\cdot; \beta)$ in \eqref{sub-model}, the Riesz representation theorem implies that $G'_{0,f}(0)$ is a linear functional of $f$. 
Consequently, the efficient influence function $\phi'_{0,f}(\cdot;0)$ is also linear in $f$.
Therefore, for any $f = \sum a_j h_j$, we have
\begin{align*}
\phi'_{P_0,f}(z; 0) = \sum_{j=1}^J a_j\phi'_{P_0,h_j}(z; 0).
\end{align*} 
Now, let $\boldsymbol{\Phi}$ be an $n \times J$ matrix with element $(i,j)$ given by $\phi'_{\hat{P}, h_j}(Z_i; 0)$, and let $\boldsymbol{\xi}_m$ be an $n$-dimensional vector of Rademacher random variables, as in Sections \ref{sec:bs-test} and \ref{sec:bs-interval}.
Similarly as $\tilde{\Psi}_n$, the multiplier bootstrap test statistics $T_{n,m}^\xi$ in \eqref{mult-bs} can be expressed as
\begin{align}
2nT^\xi_{n,m} &= \max_{\mathbf{a}} \left\{ \mathbf{a}^\top\boldsymbol{\Phi}^\top \left[\text{diag}(\boldsymbol{\xi}_m) \right]^2 \boldsymbol{\Phi} \mathbf{a}: \mathbf{a}^\top \mathbf{L} \mathbf{a} \leq \lambda, \mathbf{a}^\top \mathbf{H}_1 \mathbf{a} \leq 1 \right\}. \label{comp-bs1}
\end{align}
The optimization problem in \eqref{comp-bs1} is also a QCQP and can solved efficiently.
Finally, $U_{n,m}^\xi$ can be written as
\begin{align*}
U_{n,m}^\xi = \left|\sum_{i=1}^n\xi_{i,m}\left\{\phi_{\hat{P}_n,f}(Z_i; 0) - \tilde{a}_{n,j}\sum_{j=1}^J\phi_{\hat{P}_n,h_j}\left(Z_i; \frac{\mathbf{H}_2^\top\tilde{\mathbf{a}}_n}{\tilde{\mathbf{a}}_n^\top \mathbf{H}_1\tilde{\mathbf{a}}_n}\right) \right\} \right|,
\end{align*} 
where $\tilde{\mathbf{a}}_n = (\tilde{a}_{1, n}, \ldots, \tilde{a}_{J,n})$ is a solution to \eqref{comp-iif}.

\section{Illustration: Inference in a Nonparametric Regression Model} \label{sec:illustration1}

In this section, we apply our framework to perform inference on the non-negative dissimilarity measured described in Example 1.
In the Supplementary Materials, we describe our framework can be used to construct a test of stochastic dependence, following the setting described in Example 2.

In this problem, we are tasked with assessing whether a subset of a collection of predictor variables is needed for attaining an optimal prediction function.
As in Section \ref{sec:examples}, our data take the form $Z = (W, X, Y)$, where $Y$ is a real-valued outcome variable, and $X$ is the predictor vector of interest, and $W$ is a vector of covariates.
We denote by $\mu_{P,Y}:w \mapsto E_P[Y|W = w]$ the conditional mean of the outcome given the covariates.
Our objectives are to assess whether there exists a function that depends on both $X$ and $W$ which predicts $Y$ better than $\mu_{P_0,Y}(W)$, and to measure the best achievable improvement in predictive performance by any function in a large class.

We specify the parametric sub-model $\theta^*_{P,f}$ as
\begin{align*}
\theta^*_{P,f}(z; \beta) = \mu_{P,Y}(w) + \beta f(w,x).
\end{align*}
The goodness-of-fit of any candidate in the sub-model is defined as
\begin{align*}
G_{P,f}(\beta) := \int \left\{y - \mu_{P,Y}(w) - \beta f(w,x) \right\}^2 dP(z),
\end{align*}
and the first and second derivatives are given by
\begin{align*}
&G'_{P,f}(\beta) = -2\int f(w,x)\left\{y - \mu_{Y,P}(w) - \beta f(w,x) \right\} dP(z),
\\
&G''_{P,f}(\beta) = 2\int f^2(w,x)dP(z).
\end{align*}

As we noted in Section \ref{sec:framework}, knowledge of the efficient influence function of $G_{P,f}(\beta)$ is helpful for constructing an asymptotically linear estimator thereof.
Additionally, we require the derivative of the efficient influence function to exist.
Let $\mu_{f,P}(w) = E[f(W,X)|W = w]$ represent the conditional mean of $f(W,X)$ given $W$.
The form of the efficient influence function and its derivative are provided in the following lemma.
\begin{lemma}
The efficient influence function of $G_{P,f}(\beta) =  \int \left\{y - \mu_{Y,P}(w) - \beta f(w,x) \right\}^2 dP(z)$ is given by
\begin{align*}
\phi_{P, f}(\cdot;\beta): (w,x,y) \mapsto \left\{y - \mu_{P,Y}(w) - \beta f(w,x)\right\}^2 + 2\beta \left\{y - \mu_{Y,P}(w)\right\} \mu_{f,P}(w) - G_{P,f}(\beta).
\end{align*}
\end{lemma}
\noindent It is also easy to see that the efficient influence function is twice differentiable in $\beta$.
The evaluation of its first and second derivatives at $\beta = 0$ are given by
\begin{align*}
&\phi'_{P,f}(\cdot; 0) (w,x,y)\mapsto  -2\left\{y - \mu_{P,Y}(w)\right\}\left\{f(x,w) - \mu_{P,f}(w) \right\} - G'_{P,f}(0),
\\
&\phi''_{P,f}(\cdot; 0) (w,x,y)\mapsto  -2f^2(w,x)- G''_{P,f}(0).
\end{align*}

From Lemma 1, we can see that $\{G_{0,f}(\beta): f \in \mathcal{F}, 
\beta \in \mathcal{B}\}$ and $\{\phi_{P_0,f}(\cdot; \beta): f \in \mathcal{F}, \beta \in \mathcal{B}\}$ depend on the  nuisance parameters $\mu_{Y,P_0}$ and $\{\mu_{P_0,f}: f \in \mathcal{F}\}$.
One can obtain nonparametric estimators $\mu_{n, Y}$ and $\{\mu_{n,f}: f \in \mathcal{F}\}$ for the nuisance using any in a wide variety of flexible data-adaptive regression procedures, including kernel ridge regression \citep{wahba1990spline},  neural networks \citep{barron1989statistical}, the highly adaptive lasso \citep{benkeser2016highly}, or the Super Learner \citep{van2007super}.
In our implementation, we use kernel ridge regression, in large part because it is computationally efficient.

It may at first seem computationally difficult to estimate the conditional mean of $f(X,W)$ given $W$ for all $f$.
However, because we have assumed that $f$ can be represented as a linear combination of basis functions $h_1, h_2, \ldots$, we can perform this calculation without too much trouble.
For $f = \sum_{j=1}^J a_j h_j$, we have the representation $\mu_{P,f} = \sum_{j = 1}^J a_j \mu_{P,h_j}$.
Therefore, one can obtain estimators $\mu_{n,h_j}$ for $\mu_{P_0,h_j}$ for $j = 1, \ldots, J$ and then estimate $\mu_{P_0,f}$ as $\mu_{n,f} = \sum_{j=1}^J a_j\mu_{n,h_j}$.

Consider the following initial plug-in estimator for $G_{0,f}(\beta)$:
\begin{align*}
G_{n,f}(\beta) = \frac{1}{n}\sum_{i=1}^n \left\{Y_i - \mu_{n,Y}(W_i) - \beta f(W_i, X_i) \right\}^2,
\end{align*}
and consider the efficient influence function estimator
\begin{align*}
\phi_{n, f}(w,x,y;\beta) = \left\{y - \mu_{n,Y}(w) - \beta f(w,x)\right\}^2 + 2\beta \left\{y - \mu_{n,Y}(w)\right\} \mu_{n,f}(w) - G_{n,f}(\beta).
\end{align*}
The initial estimator for $G_{0,f}(\beta)$ is biased, so a corrected estimator is needed so that one can perform inference.
We can use the following one-step bias-corrected estimator:
\begin{align}
\tilde{G}_{n,f}(\beta) &= G_{n,f}(\beta) + \frac{1}{n}\sum_{i=1}^n \phi_{n,f}(Z_i; \beta) \nonumber
\\
&= \frac{1}{n}\sum_{i=1}^n \left[ \left\{Y_i - \mu_{n,Y}(W_i) - \beta f(W_i, X_i) \right\}^2 +  2 \beta \left\{ Y_i - \mu_{n,Y}(W_i) \right\} \mu_{n,f}(W_i) \right].
\label{onestep-npregress}
\end{align}
The following lemma provides conditions under which the one-step estimator satisfies Assumptions B1 through B4.
\begin{lemma}
Suppose that the nuisance estimators satisfy the rate conditions
\begin{align*}
&\left[\int \left\{\mu_{Y, P_0}(w) - \mu_{Y, n}(w) \right\}^2 dP_0(w)\right]^{1/2}  = o_P(n^{-1/4}),
\\
\sup_{f\in\mathcal{F}}&\int\left|\left\{\mu_{P_0, f}(w) - \mu_{n,f}(w) \right\}\left\{\mu_{Y, P_0}(w) - \mu_{Y, n}(w) \right\} dP_0(w) \right|  = o_P(n^{-1/2}).
\end{align*}
Suppose also that there exist $P_0$-Donsker classes $\Phi$, $\Phi'$ and a $P_0$-Glivenko-Cantelli class $\Phi''$ such that, with probability tending to one, each of the following holds:
\begin{align*}
&\{\phi_{n,f}(\cdot;\beta) - \phi_{P_0,f}(\cdot; \beta): f \in \mathcal{F}, \beta \in \mathcal{B}\} \subset \Phi,
\\
&\{\phi'_{n,f}(\cdot;0) - \phi'_{P_0,f}(\cdot; 0): f \in \mathcal{F}\} \subset \Phi',
\\
&\{\phi''_{n,f}(\cdot;0) - \phi''_{P_0,f}(\cdot; 0): f \in \mathcal{F}\} \subset \Phi''.
\end{align*}
 Then the one-step estimator $\tilde{G}_{n,f}$ in \eqref{onestep-npregress} satisfies Assumptions B1-B4.
\end{lemma}
\noindent The condition on the convergence rates of the nuisance estimators is standard and holds when all nuisance estimators are $n^{1/4}$-rate convergent.
This is rate is attained by many nonparametric regression estimators under weak structural assumptions on the true conditional mean functions, so the condition is fairly mild.

We conclude with a brief comment about computation.
We observe that $\tilde{G}_{n,f}(\beta)$ is a quadratic function of $\beta$, so the implementation scheme described in Section \ref{sec:comp} can be applied in this example.

\section{Simulation Study} \label{sec:simulations}

In this section, we assess the empirical performance of our proposal in our simulation study.
In this example, we again consider the nonparametric regression task discussed in Section \ref{sec:illustration1}.

We generate synthetic data sets as follows.
First, we generate independent $3$-dimensional random vectors $A_1, \ldots, A_n$ from a Gaussian distribution with mean zero and covariance
\begin{align*}
\mathbf{V} = 
\begin{pmatrix}
1 & .5 & .5 
\\
.5 & 1 & .5
\\
.5 & .5 & 1
\end{pmatrix}.
\end{align*}
We then define the predictor vector as $X_i = (2\gamma(A_{i,1}) - 1, 2 \gamma(A_{i,2}) - 1, 2 \gamma(A_{i,3}) - 1)$, where $\gamma$ is the standard normal distribution function.
We generate the outcome $Y$ according to the model
\begin{align*}
Y_i = \sin(\pi X_{i,1}) - 2\left( X_{i,2} - \frac{1}{2}\right)^2 + \mathds{1}(X_{i,2} > 0)\exp(X_{i,1})+ \epsilon_i,
\end{align*}
where the white noise $\epsilon_i$ is a continuous uniform $[-6, 6]$ random variable, drawn independently of $X_i$.
Our objective is to determine which of the elements of the predictor $X$ are conditionally associated with the outcome $Y$, given the other elements.
Clearly, $Y$ is conditionally dependent on the first and second elements, and not the third.

Our target of inference is the improvement in fit estimand defined in Example 1.
More specifically, for each predictor, we estimate the improvement in fit comparing a flexible regression model containing prediction functions that may depend on all predictors, with a model only containing functions that do not depend on the predictor of interest.
We estimate all nuisance parameters described in Section \ref{sec:illustration1} using kernel ridge regression.
We construct $\mathcal{F}$ using the reproducing kernel Hilbert space corresponding to the Gaussian kernel, and we consider two choices for the smoothness parameter.
First, we use an oracle apporach, which sets $\mathcal{F}$ as the smoothest class containing a function that is  proportional to the difference between the full conditional mean of $Y$ given all predictors $X$, and the conditional mean of $Y$ given predictors that are not of interest.
We recall from our discussion in Section \ref{sec:examples} that the difference between conditional means is the function that maximizes the improvement in fit.
Second, we consider a data-adaptive procedure, where the smoothness parameter is selected using cross-validation, and no sample-splitting is performed.

We compare our method with a sample splitting approach proposed by \cite{williamson2021general}.
They propose to separate the dataset into two independent partitions -- one of which is used to estimate the optimal goodness-of-fit over the full model, and the second of which is used to estimate  the optimal goodness-of-fit over the reduced model.
Then inference on $\Psi_0$ can be performed using a two-sample Wald test, and Wald-type intervals can similarly be constructed.
We expect our approach to offer an improvement because we do not require sample-splitting for valid inference.
We also expect our proposal to benefit from achieving fast $n$-rate convergence rate at the boundary, compared with the sample-splitting approach, which is only $n^{1/2}$-rate convergent.

We generate 1000 synthetic data sets under the data-generating process described above for $n \in \{100, 200, 400, 800, 1600, 3200\}$.
We compare all methods under consideration in terms mean squared error, type-1 error control under the null, power under the alternative, confidence interval coverage, and average confidence interval width.

\begin{figure}[!h]
\center
\includegraphics[scale=.75]{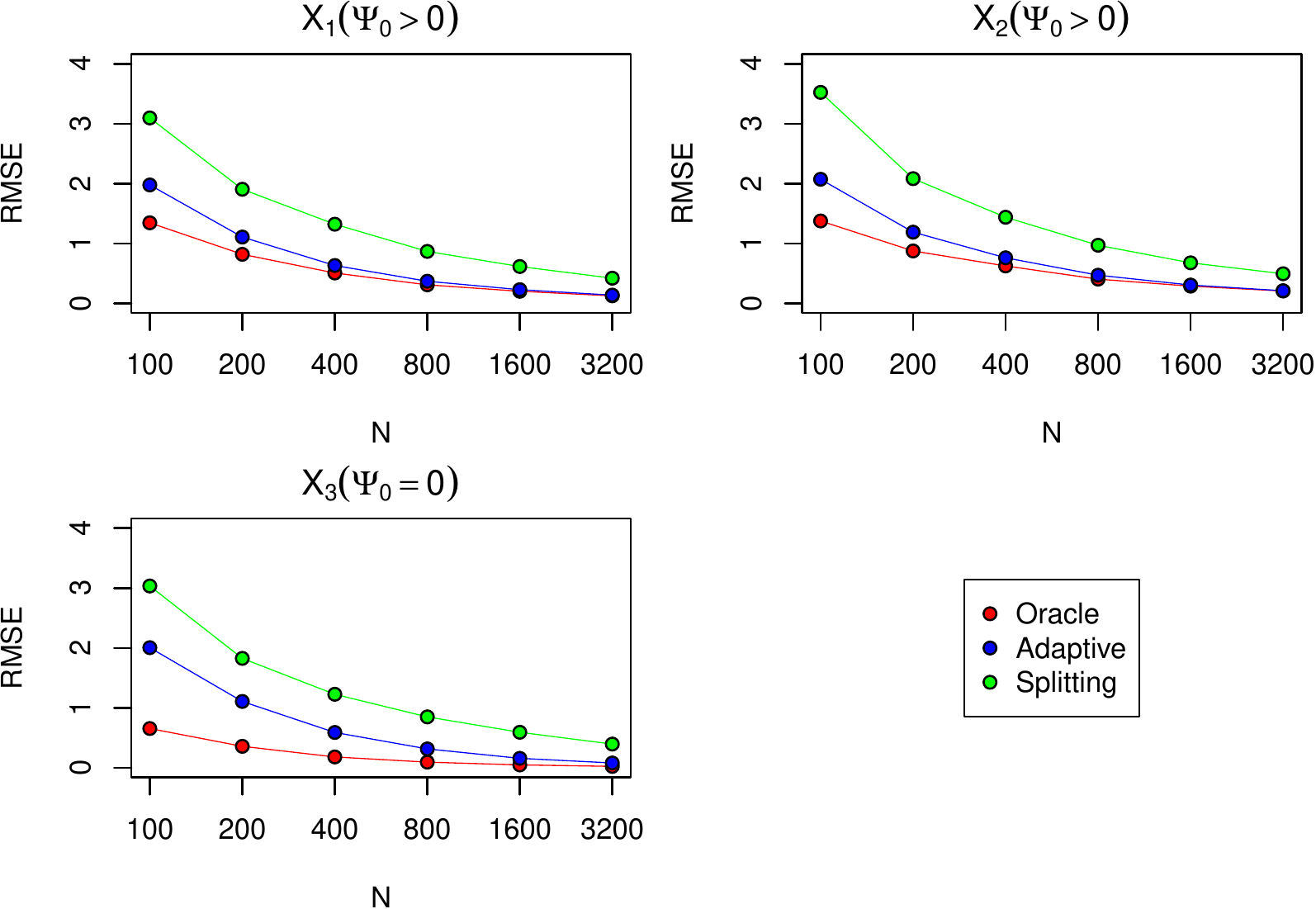}
\caption{Monte Carlo estimates of the root mean squared error for improvement in fit estimators in our simulation study.}
\label{fig:rmse}
\end{figure}

Figure \ref{fig:rmse} shows the root mean squared error for each proposed improvement in fit estimator as a function of the sample size.
We find that, while the root mean squared error of each estimator approaches zero, our proposed estimators converge much more quickly than the sample splitting estimator, with the oracle version of our approach performing best.

\begin{figure}[!h]
\center
\includegraphics[scale=.75]{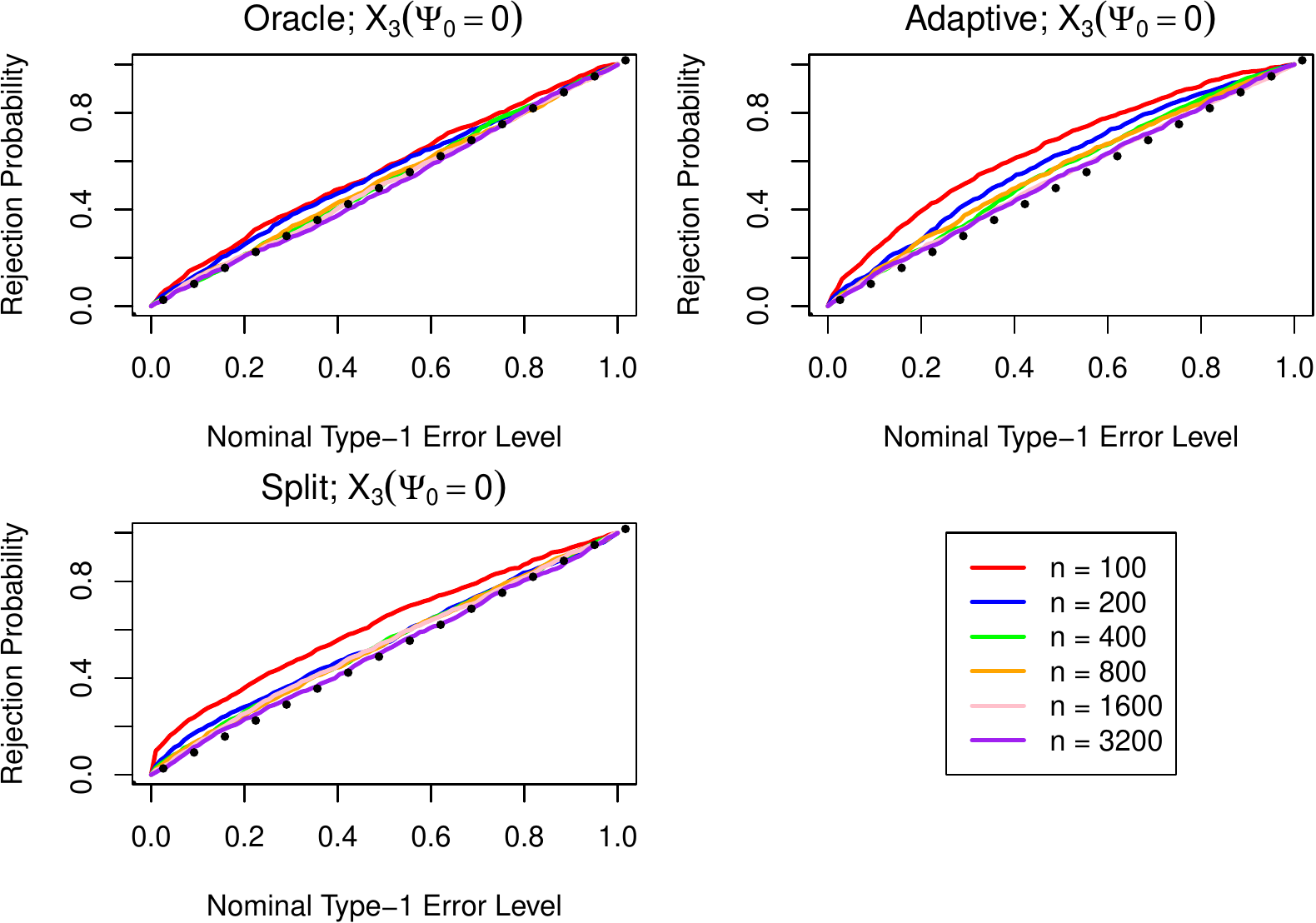}
\caption{Monte Carlo estimates of the rejection probability as a function of the nominal type-1 error level, under the null of no improvement in fit in our simulation study.}
\label{fig:type1}
\end{figure}

\begin{figure}[!h]
\center
\includegraphics[scale=.85]{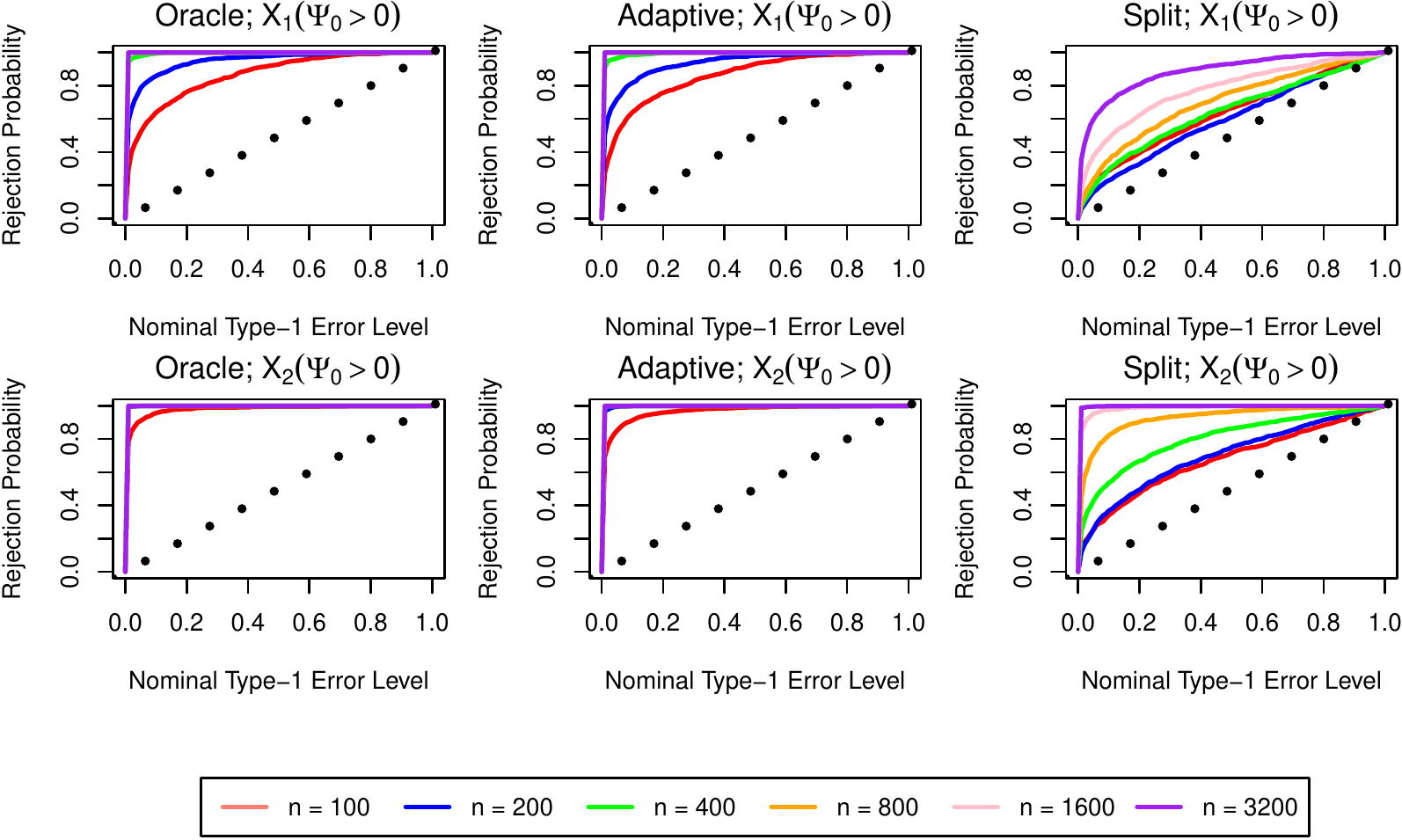}
\caption{Monte Carlo estimates of the rejection probability as a function of the nominal type-1 error level, under the alternative of positive improvement in fit in our simulation study.}
\label{fig:power}
\end{figure}

In Figures \ref{fig:type1} and \ref{fig:power} we plot the rejection probability for a test of the null of no improvement in fit as a function of the nominal type-1 error level.
We find that all tests considered achieve asymptotic type-1 error control in the setting where $\Psi_0 = 0$, though we acknowledge there is type-1 error inflation when the sample size is small.
We find that our proposed tests are well-powered against the null, both outperforming the sample-splitting estimator.

\begin{figure}[!h]
\center
\includegraphics[scale=.85]{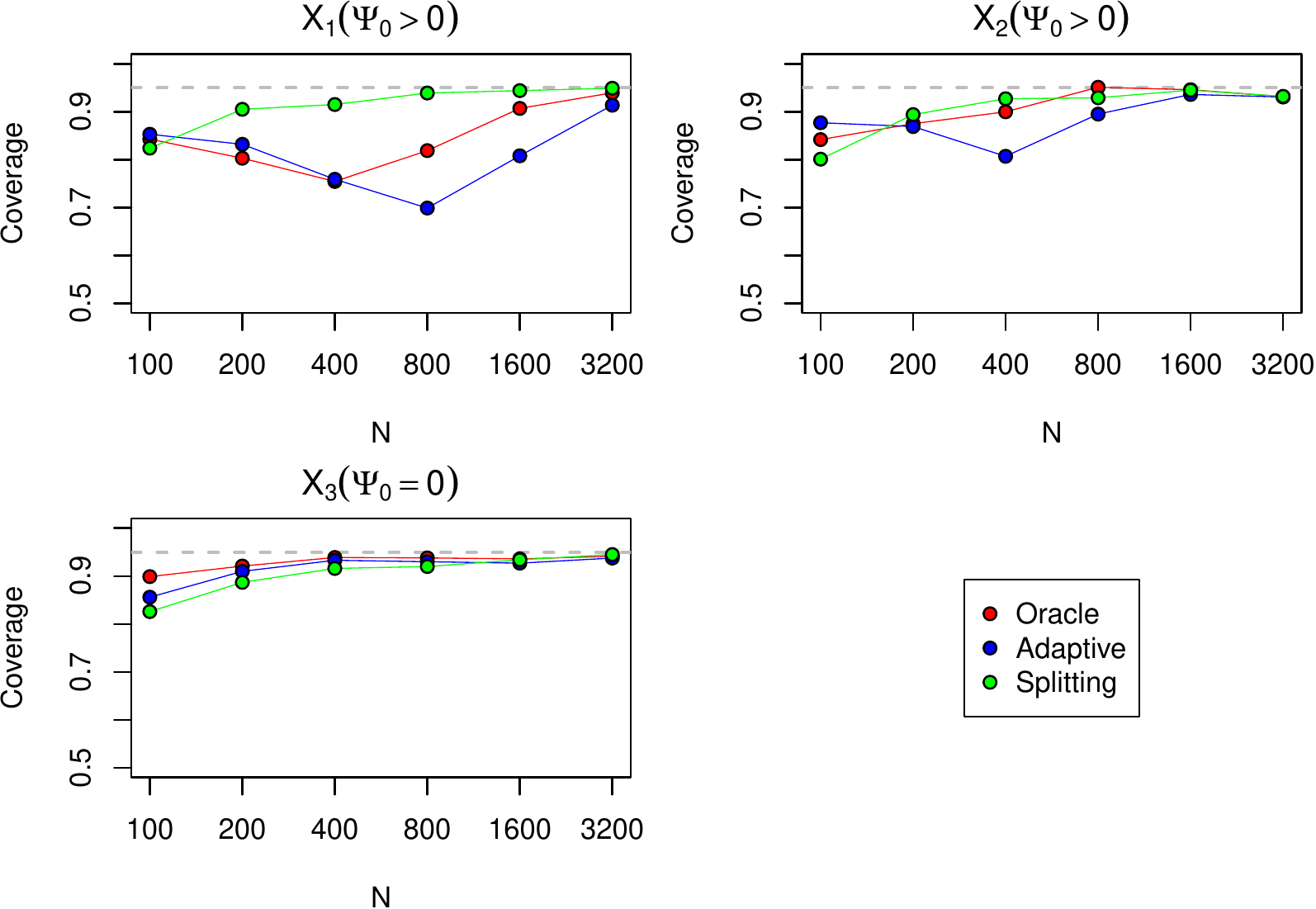}
\caption{Monte Carlo estimates of coverage probability of 95\% confidence intervals for the improvement in fit in our simulation study.}
\label{fig:coverage}
\end{figure}

\begin{figure}[!h]
\center
\includegraphics[scale=.85]{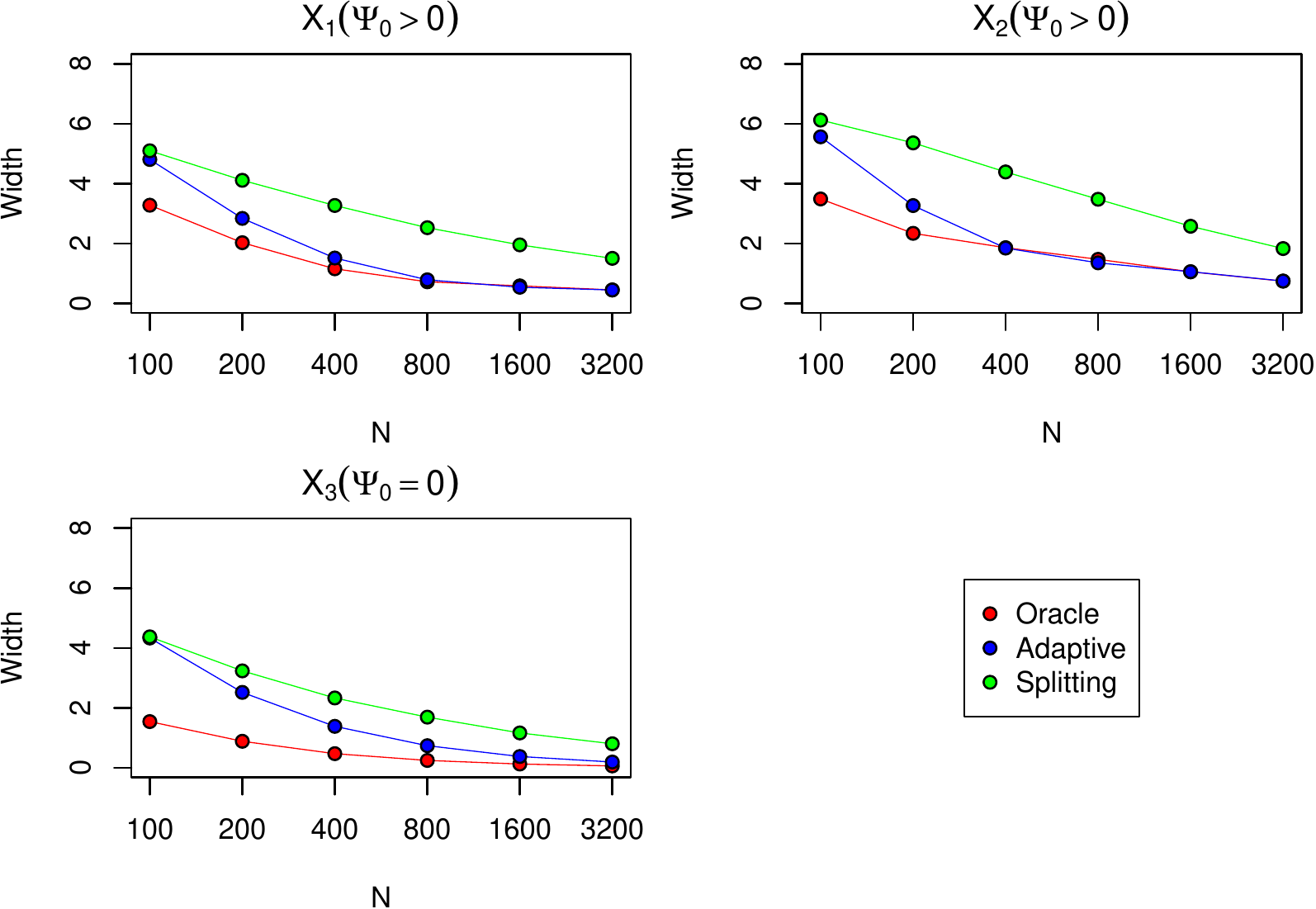}
\caption{Monte Carlo estimates of average width of 95\% confidence intervals for the improvement in our simulation study.}
\label{fig:width}
\end{figure}

In Figures \ref{fig:coverage}  and \ref{fig:width} we plot coverage probability and average width of 95\% confidence intervals as a function of the sample size.
We find that all approaches considered achieve nominal coverage as the sample size grows, though the there is a tendency for our proposed intervals to exhibit undercoverage when the sample size is small.
Our proposed estimator with oracle selection of $\mathcal{F}$ achieves the lowest average width, followed by the adaptive approach, and the sample-splitting approach.


\section{Discussion} \label{sec:discussion}

In this work, we have presented a general framework for inference on non-negative dissimilarity measures.
Our proposed methodology has wide-ranging utility.
As examples, we described how this framework can be applied to perform rate-optimal inference on statistical functionals arising in nonparametric regression and graphical modeling problems.
Our framework can also be useful in other settings, such as causal inference problems.
For instance, some statistical functionals that have been used for studying treatment heterogeneity \citep[see, e.g.,][]{levy2021fundamental, hines2021parameterising, sanchez2023robust} have the representation described in Section \ref{sec:prelim}, so one can perform inference using our general approach.

Our work has some notable limitations that we plan to address in future research.
While  our proposal for inference on the improvement in fit enjoys good behavior in a large sample setting, we observed in our simulatoin study that it may have undesirable small sample properties, such mild type-1 error inflation or poor coverage.
Additionally, our estimator suffers a loss in precision when we select of $\mathcal{F}$ in a data-adaptive manner.
In future work, we plan to investigate whether the performance can be improved using, e.g., small-sample adjustments or improved data-adaptive methods for tuning parameter selection.
Additionally, because our results assume smoothness of the goodness-of-fit functional, it is not clear whether our results can be directly applied to perform inference on estimands such as $L_1$ distances.
It is of interest to develop a more flexible inferential strategy that relaxes this assumption.
Our methodology also places complexity constraints on nuisance parameter estimators, which prohibits us from using estimators such as gradient boosted trees \citep{friedman2002stochastic}.
It is of interest to develop cross-fitted versions of our improvement in fit estimator and multiplier bootstrap strategy that relax this assumption \citep{zheng2011cross, chernozhukov2018double}.

There also remain several open theoretical and methodological questions.
For instance, while we have established $n-$rate consistency of our proposed improvement in fit estimator, it is unclear whether our test of the null of no improvement in fit is optimal in any sense.
It would be important to characterize the power of our test and to determine whether there exists a more powerful test.
Additionally, it is of interest to understand how specification of the sub-model $\theta^*_{P,f}$ affects our procedure's performance.
It is possible that there are many ways for one to construct a sub-model while still obtaining valid inference on $\Psi_0$.
It is not clear how this choice affects the estimator or whether there is an optimal choice.
We expect that, in practice, this choice will need to be made in consideration of theoretical properties, such as power, and more practical concerns, such as ease of implementation and computational efficiency.
\newpage

\newpage

\bibliography{nplrt.bib}

\newpage

\section*{\centering Supplementary Materials}

\section*{S1 Implementation for Non-quadratic Objectives}

Here, we propose a general method for computing the improvement in fit estimator.
As noted in Section \ref{sec:comp}, computation can be challenging because when we use the specification of $\mathcal{F}$ in \eqref{implementation-class}, the optimzation problem in \eqref{iif-estimator} is possibly non-convex.
Moreover, a closed form expression for $\tilde{\beta}_{n,f}$ may not be available, further complicating the problem.

As in Section \ref{sec:comp}, we focus on the setting where the complexity measure $\Gamma$ is available in quadratic form, satisfying $\Gamma(f) = \mathbf{a}^\top \mathbf{L}\mathbf{a}$ for some matrix $\mathbf{L}$.
Also, we note that in general, the second derivative of the goodness-of-fit $G_{P,f}''(0)$ is available in quadratic in the coefficients as a consequence of the Riesz representation theorem. 
We assume that $\tilde{G}_{n,f}''(0)$ can be expressed as $\mathbf{a}^\top \Omega \mathbf{a}$ for some $\Omega$.

We propose a slightly-modified estimator for $\tilde{\Psi}_0$ that has nearly identical asymptotic properties to $\tilde{\Psi}_n$, but which can be easier to compute in a more general setting.
The main idea is to separate estimation of $\Psi_0$ into two parts.
First, as before, for each $f \in \mathcal{F}$ we perform a search to identify whether any candidate parameter in the sub-model is an improvement over the null in terms goodness-of-fit.  
Where we differ is that we confine our search to a small neighborhood of zero.
When there exists evidence that for some $f$, $G_{0,f}$ is not minimized at zero, we search over a larger function class to identify a candidate parameter that achieves a better fit than the null.
In contrast to the strategy proposed in Section \ref{sec:implementation}, we specify $\Theta \setminus \Theta^*$ as a class over which an optimum can more easily be identified.
In what follows, we provide details and rationale for this method.

First, Taylor's theorem implies that when $\beta_{0,f}$ resides within a neighborhood of zero, $\beta_{0,f}$ is approximately equal to $\{G''_{0,f}(0)\}^{-1} G'_{0,f}(0)$. 
Additionally, in this setting we have that
\begin{align*}
\Psi_0^* := \sup_{f \in \mathcal{F}} \frac{\left\{G'_{0,f}(0) \right\}^2}{2G''_{0,f}(0)} \approx\sup_{f \in \mathcal{F}} G_{0,f}(0) - G_{0,f}(\beta_{0,f}) = \Psi_0.
\end{align*}
Because $G_{0,f}$ is assumed to be convex for all $f$, then $G'_{0,f^*}(0) \neq 0$ for some $f^*$, implies that $\beta_{0,f^*} \neq 0$, and $\Psi_0 > 0$.
And so, one can assess whether $\Psi_0 = 0$ by checking whether $\Psi_0^* = 0$.
This is roughly equivalent to performing a search over $\mathcal{F} \times \mathcal{B}$ to identify the best fit, where $\mathcal{B}$ is taken to be a small interval containing zero.
To estimate $\Psi^*_0$, we use the estimator
\begin{align*}
\tilde{\Psi}_n^* := \sup_{f \in \mathcal{F}} \frac{\left\{\tilde{G}_{n,f}'(0) \right\}^2}{2\tilde{G}''_{n,f}(0)}.
\end{align*}
When the null holds, and the conditions of Theorem 2 hold as well, $\tilde{\Psi}^{*}_n$ has the same asymptotic representation as $\tilde{\Psi}_n$. That is,
\begin{align*}
\tilde{\Psi}^{*}_n = \sup_{f \in \mathcal{F}}\frac{1}{2 G''_{0,f}(0)} \left[ \frac{1}{n}\sum_{i=1}^n \phi'_{P_0, f}(Z_i; 0)\right]^2 + o_P(n^{-1}). 
\end{align*}

As noted in Section \ref{sec:comp}, the Riesz representation theorem implies that $G'_{0,f}(0)$ is a linear functional of $f$.
We assume that our estimator $\tilde{G}_{n,f}(0)$ is also linear, and can be expressed as $\mathbf{H}^\top\mathbf{a}$ for some $J$-dimensional vector $\mathbf{H}$.
Thus, using the specification of $\mathcal{F}$ in \eqref{implementation-class}, $\tilde{\Psi}_n^*$ can be expressed as
\begin{align*}
2\tilde{\Psi}^*_n &= \max_{\mathbf{a}} \left\{ \mathbf{a}^\top\mathbf{H} \mathbf{H}^\top \mathbf{a}: \mathbf{a}^\top \mathbf{L} \mathbf{a} \leq \lambda, \mathbf{a}^\top \Omega \mathbf{a} \leq 1 \right\}.
\end{align*}
This problem is a quadratically constrained quadratic program and can be solved efficiently.

It is possible that $\tilde{\Psi}_n^*$ is a poor approximation for $\tilde{\Psi}_n$ when the null of no improvement in fit does not hold.
Let $\pi^*_n \in (0,1)$ be a random sequence that converges to one in probability when $\Psi_0^* = 0$ holds and converges to zero in probability when $\Psi_0^* > 0$.
For instance, we can take $\pi^*_n$ as
\begin{align*}
\pi^*_n = \rho_0\left( \frac{n}{\log(n)} \tilde{\Psi}_n^* \right),
\end{align*}
where $\rho_0$ is as defined in \eqref{p-value}.
For large values of $\pi^*_n$, $\tilde{\Psi}^*_n$ can replace $\tilde{\Psi}_n$.
Otherwise, an alternative estimator may be needed.

To estimate $\Psi_0$ when the null does not hold, we consider an alternative specification of $\Theta \setminus \Theta^*$.
A main source of our difficulty with solving the optimization problem in \eqref{iif-estimator} is that the constraint $\{G''_{0,f}(0)\}^{-1}\Gamma(f) \leq \lambda$ is non-convex.
Under the null, constraining $\{G''_{0,f}(0)\}^{-1}$ is necessary, as Theorem 2 states that Assumption A4 must hold in order for our improvement in fit estimator to have a well-behaved limiting distribution.
However, this assumption is not needed for Theorem 3 to hold.
As we discussed previously in Section \ref{sec:implement-class}, for Theorem 3 to hold, we really only need to ensure that the variance of $\{\tilde{\beta}_{n,f}: f \in \mathcal{F}\}$ is well controlled.
When $G_{0,f}$ is quadratic, this can be achieved by constraining $\{G''_{0,f}(0) \}^{-1}$ and leaving $\mathcal{B}$ unconstrained, as was done previously.
When $G_{0,f}$ is non-quadratic, we can alternatively leave $\{G''_{0,f}(0)\}^{-1}$ unconstrained and carefully select the width of $\mathcal{B}$.
We instead set $\mathcal{F}$ as the function class
\begin{align*}
\mathcal{F}^*_\lambda :=  \left\{ f = \sum_{j=1}^\infty a_j h_j: \Gamma(f) \leq \lambda \right\},
\end{align*}
and we set $\mathcal{B}$ as $\mathcal{B}^*_\sigma := [-\sigma, \sigma]$ for some $\sigma > 0$.
Now, we define $\Psi_0^{**}$  as
\begin{align*}
\Psi_0^{**} := \inf_{f \in \mathcal{F}^*_\lambda} \inf_{\beta \in \mathcal{B}^*_\sigma} G_{0,f}(\beta),
\end{align*}
and we estimate $\tilde{\Psi}_0^{**}$ as
\begin{align*}
\tilde{\Psi}_n^{**} := \inf_{f \in \mathcal{F}^*_\lambda} \inf_{\beta \in \mathcal{B}^*_\sigma} \tilde{G}_{n,f}(\beta) .
\end{align*}

Calculating $\tilde{\Psi}_n^{**}$ will in many cases be much easier than calculating $\tilde{\Psi}_n$.
We can write $\tilde{\Psi}_n^{**}$ as
\begin{align*}
\tilde{\Psi}_n^{**}
= 
\inf_{f \in \mathcal{F}} \inf_{\beta \in \mathcal{B}} \tilde{G}_{n,\beta f}(1) 
= 
\min_{\mathbf{b}} \left\{ \tilde{G}_{n, \sum b_j h_j}(1): \mathbf{b}^\top \mathbf{L} \mathbf{b} \leq \sigma \lambda \right\},
\end{align*}
where $\mathbf{b} = (b_1, \ldots, b_J)$ is a $J$-dimensional vector.
This optimization problem has a only single convex constraint, so the problem will be convex when the objective function is also convex.
This can greatly simplify computation.

Of course, $\tilde{\Psi}_n^{**}$ and $\tilde{\Psi}_n$ potentially can achieve different limiting distributions when $\Psi > 0$.
This is because $\Psi_0^{**}$ and $\Psi_0$ are defined as optima over different function classes.
While the two values are expected to be similar, they will not necessarily be equal.
Nonetheless, one can still apply Theorem 3 to establish weak convergence of $\tilde{\Psi}_n^{**}$ to a Gaussian distribution, as long as $\mathcal{F} \times \mathcal{B}$ is not too large.

Finally, we combine $\tilde{\Psi}_n^*$ and $\tilde{\Psi}_n^{**}$ to obtain a single estimator for $\Psi_0$.
We define our estimator as
\begin{align*}
\check{\Psi}_n = \pi_n^* \tilde{\Psi}_n^* + (1 - \pi_n^*) \tilde{\Psi}_n^{**}.
\end{align*}
Because $\pi^*_n$ tends to one when the null holds and approaches zero when the null fails, $\check{\Psi}_n$ has approximately the same asymptotic behavior as $\tilde{\Psi}_n$.
That is $\check{\Psi}_n$ behaves like the supremum of an empirical process under the null, and like a sample average otherwise.
Therefore, the multiplier bootstrap tests described in Sections \ref{sec:bs-test} and \ref{sec:bs-interval} remain valid, and a similar strategy for implementation as described in Section \ref{sec:comp} can be used.

\section*{S2 Illustration: Nonparametric Assessment of Stochastic Dependence}

In this Section, we briefly discuss inference in Example 2 from Section \ref{sec:examples}, where we are interested in assessing whether a pair of random variables is independent.
Our data takes the form $Z = (X, Y)$, where $X$ and $Y$ are one-dimensional random variables.

We assess dependence by comparing the expectation of the log of the product of the marginal densities of $X$ and $Y$ with the expectation of the logarithm of an approximation for the joint density.
Let $q_{P,X}$ and $q_{P,Y}$ denote the marginal densities of $X$ and $Y$ under $P$, and let $\theta^*_P = \log q_{P,X} q_{P,Y}$ denote the log of the product of the marginal densities. 
We use the following sub-model to approximate the logarithm of the joint density:
\begin{align*}
 \theta^*_{P,f}:(x,y) \mapsto  \theta^*_P(x,y) + \beta f(x,y) - \log \int \exp( \theta^*_P(x,y) + \beta f(x,y)) d\mu(x,y).
\end{align*}
With a straightforward calculation, it can be verified that \eqref{submod-prop-1} is satisfied, and moreover, we can see that $\theta^*_{P,f}$ is a valid candidate log density, as for any $f$, $\int \exp \theta^*_{P,f}(z; \beta) d\mu(z) = 1$.

With the above specification for the parametric sub-model, the goodness-of-fit takes the form
\begin{align*}
G_{P,f}(\beta) = E_P[-\log q_{P,X}(X) - \log q_{P,Y}(Y) - \beta f(X,Y)] + \log \left(E_{P_X}E_{P_Y}[\exp(\beta f(X,Y))] \right),
\end{align*}
where $E_{P_X}$ and $E_{P_Y}$ denote the marginal expectations under $P$, with respect to $X$ and $Y$, respectively.
We now observe that along any submodel, the difference in goodness-of-fit comparing a given candidate parameter with the null is given by
\begin{align*}
\psi_{P,f}(\beta) := G_{P,f}(\beta) - G_{P,f}(0) = -\beta E_P[ f(X,Y)] + \log \left(E_{P_X}E_{P_Y}[\exp(\beta f(X,Y))] \right),
\end{align*}
and this expression does not depend on the marginal densities $q_{P,X}$ and $q_{P,Y}$.
Thus, estimation of the marginal densities is not needed.

The derivatives $G'_{P,f}(0)$ and $G''_{P,f}(0)$ are given by
\begin{align*}
&G'_{P,f}(0) = \frac{d}{d\beta} \psi_{P,f}(\beta) \bigg \vert_{\beta = 0} = -E_P[ f(X,Y)] + E_{P_X}E_{P_Y}[f(X,Y)],
\\
&G''_{P,f}(0) = \frac{d^2}{d\beta^2} \psi_{P,f}(\beta) \bigg \vert_{\beta = 0} =  E_{P_X}E_{P_Y}[f^2(X,Y)] - \left\{E_{P_X}E_{P_Y}[f(X,Y)] \right\}^2.
\end{align*}
One can interpret $G'_{P,f}(0)$ as the difference between the true mean of $f(X,Y)$ under $P$ and the value the mean would hypothetically take if $X$ and $Y$ were independent.
The second derivative $G''_{P,f}(0)$ represents the variance of $f(X,Y)$ under the assumption that $X$ and $Y$ are independent.
 
Because $\psi_{P,f}(\beta)$ does not depend on $P$ through any nuisance parameters that are not pathwise differentiable, it is expected that a plug-in estimator, which is defined as a functional of the cumulative distribution function, would be asymptotically linear, and no sophisticated methods for bias correction should be needed.
We use the following plug-in estimator for $\psi_{P_0,f}(\beta)$:
\begin{align*}
\tilde{\psi}_{n,f}(\beta) =  \frac{-\beta}{n}\sum_{i=1}^n f(X_i, Y_i) + 
\log\left\{ \frac{1}{n^2} \sum_{i=1}^n \sum_{j=1}^n \exp(\beta f(X_i, Y_j)) \right\}.
\end{align*}
By an application of the functional delta method, one can show that the plug-in estimator is asymptotically linear with influence function
\begin{align*}
\phi_{P,f}(z; \beta) - \phi_{P,f}(z; 0) = &\left\{\frac{E_{P_Y}[\exp(\beta f(x,Y))] + E_{P_X}[\exp(\beta f(Y,x))]}{E_{P_X}[E_{P_Y}[\exp \beta f(X,Y)]]} - \beta f(x,y)\right\} - \left\{2 - E_P[\beta f(X,Y)] \right\}.
\end{align*}
Similarly, we estimate $G'_{0,f}(0)$ as
\begin{align*}
\tilde{G}'_{n,f}(0) = \frac{d}{d\beta} \tilde{\psi}_{n,f}(\beta) \bigg\vert_{\beta = 0} = \frac{-1}{n}\sum_{i=1}^n f(X_i, Y_i) + \frac{1}{n^2}\sum_{i_1 = 1}^n \sum_{i_2 = 1}^n f(X_{i_1}, Y_{i_2}),
\end{align*}
and $\tilde{G}_{n,f}(0)$ is asymptotically linear with efficient influence function
\begin{align*}
\phi'_{P,f}(z; 0) = -f(x,y) + E_{P}[f(X,Y)] + \left\{ E_{P_Y}\left[ f(x,Y) \right] + E_{P_X}\left[f(X,y)\right]  - 2 E_{P_X}[E_{P_Y}[f(X,Y)]]\right\}.
\end{align*}
We estimate the second derivative $G''_{0,f}(0)$ as
\begin{align*}
\tilde{G}'_{n,f}(0) = \frac{d^2}{d\beta^2} \tilde{\psi}_{n,f}(\beta) \bigg\vert_{\beta = 0} = \frac{1}{n^2}\sum_{i_1 = 1}^n \sum_{i_2 = 1}^n f^2(X_{i_1}, Y_{i_2}) - \left\{\frac{1}{n^2}\sum_{i_1 = 1}^n \sum_{i_2 = 1}^n f(X_{i_1}, Y_{i_2}) \right\}^2.
\end{align*}

In this setting, a closed form solution for $\tilde{\beta}_{n,f}$ is not available, and if one uses the specification for $\mathcal{F}$ in \eqref{implementation-class}, the problem
\begin{align*}
\inf_{f \in \mathcal{F}, \beta \in \mathcal{B}} \tilde{\psi}_{n,f}(\beta)
\end{align*}
is difficult to solve.
As an alternative, we recommend using the more general implementation strategy presented in the Supplementary Materials Section S1.

\section*{S3 Proofs of Theoretical Results}

\subsubsection*{Proof of Theorem 1}
We have by definition that $\left\{G_{0,f}(\tilde{\beta}_{n,f}) - G_{0,f}(\beta_{0,f}) \right\}$ and $\left\{\tilde{G}_{n,f}(\beta_{0,f}) - \tilde{G}_{n,f}\left(\tilde{\beta}_{n,f}\right)  \right\}$ are non-negative.
Under Assumption B2, we can write
\begin{align*}
0 &\leq \sup_{f \in \mathcal{F}}\left\{G_{0,f}(\tilde{\beta}_{n,f}) - G_{0,f}(\beta_{0,f}) \right\} - \left\{\tilde{G}_{n,f}\left(\tilde{\beta}_{n,f}\right) - \tilde{G}_{n,f}(\beta_{0,f}) \right\}
\\
 &= \sup_{f \in \mathcal{F}}\frac{1}{n}\sum_{i=1}^n \left\{\phi_{P_0,f}(Z_i; \tilde{\beta}_{n,f}) - \phi_{P_0,f}(Z_i; \beta_{0,f})  \right\} + o_P(n^{-1/2})
 \\
&\leq \sup_{f \in \mathcal{F}, \beta \in \mathcal{B}}\frac{1}{n}\sum_{i=1}^n \left\{\phi_{P_0,f}(Z_i; \beta) - \phi_{P_0,f}(Z_i; \beta_{0,f})  \right\} + o_P(n^{-1/2}).
\end{align*}
Thus from the fact that $\{\phi_{P_0, f}(\cdot; \beta): f \in \mathcal{F}, \beta \in \mathcal{B}\}$ is $P_0$-Donsker, we have that $G_{0,f}(\tilde{\beta}_{n,f}) - G_{0,f}(\beta_{0,f}) = O_P(n^{-1/2})$.
Assumption A3 implies that $\sup_{f \in \mathcal{F}} |\tilde{\beta}_{n,f} - \beta_{0,f}| = o_P(1)$.

Now, because we have that $\tilde{\beta}_{n,f}$ satisfies $\tilde{G}'_{n,f}(\tilde{\beta}_{n,f}) = o_P(n^{-1/2})$, Taylor's theorem implies
\begin{align*}
\tilde{G}'_{n,f}(\tilde{\beta}_{n,f}) =  \tilde{G}'_{n,f}(\beta_{0,f}) + G''_{n,f}(\bar{\beta}_{n,f})(\tilde{\beta}_{n,f} - \beta_{0,f}) = 0,
\end{align*}
for some $\bar{\beta}_{n,f}$ that satisfies $|\bar{\beta}_{n,f} - \beta_{0,f}| \leq |\tilde{\beta}_{n,f} - \beta_{0,f}|$.
By rearranging terms and invoking Assumption B4, the estimation error for $\tilde{\beta}_{n,f}$ can be expressed as
\begin{align*}
\tilde{\beta}_{n,f} - \beta_{0,f} = -\left\{G''_{0,f}(\bar{\beta}_{n,f}) + o_P(1) \right\}^{-1}\tilde{G}'_{n,f}(\beta_{0,f}).
\end{align*}
Because $G'_{0,f}(\beta_{0,f}) = 0$, Assumption B2 implies that
\begin{align*}
\tilde{\beta}_{n,f} - \beta_{0,f} = -\left\{G''_{0,f}(\bar{\beta}_{n,f}) + r''_{n,f}(\bar{\beta}_{n,f}) \right\}^{-1}\left\{\frac{1}{n} \sum_{i=1}^n \phi'_{0,f}(Z_i; \beta_{0,f}) + r'_{n,f}(\beta_{0,f}) \right\}.
\end{align*}
Now, because $\{\tilde{\beta}_{n,f}: f \in \mathcal{F}\}$ is uniformly consistent for $\{\beta_{0,f}: f \in \mathcal{F}\}$, the continuous mapping theorem and Assumptions A4 and B4 allow us to replace $\left\{G''_{0,f}(\bar{\beta}_{n,f}) + r''_{n,f}(\bar{\beta}_{n,f}) \right\}^{-1}$ with $\left\{G''_{0,f}(\beta_{0,f}) \right\}^{-1}$ in the above display.
Thus, we have
\begin{align*}
\tilde{\beta}_{n,f} - \beta_{0,f} = \frac{-1}{G''_{0,f}(\beta_{0,f})} \left\{\frac{1}{n}\sum_{i=1}^n \phi_{P_0,f}(Z_i; \beta_{0,f}) \right\} + o_P(n^{-1/2}),
\end{align*}
as claimed.
The weak convergence result follows as an immediate consequence of the Donsker Assumption A5.

\subsubsection*{Proof of Theorem 2}

This result follows directly from an application of the continuous mapping theorem.

\subsubsection*{Proof of Theorem 3}

Following from our discussion from Section \ref{sec:asymp}, it suffices to show that $ G_{n,f}(\tilde{\beta}_{n,f}) - G_{n,f_0}(\beta_{0,f_0}) = o_P(n^{-1/2})$. 
First, we write
\begin{align*}
\sup_{f \in \mathcal{F}} G_{n,f}(\tilde{\beta}_{n,f}) - G_{n,f_0}(\beta_{0,f_0}) 
&\leq
\left\{G_{n,f_n}(\tilde{\beta}_{n,f}) - G_{n,f_0}(\beta_{0, f_0})\right\} - \left\{ G_{0,f_n}(\tilde{\beta}_{n,f_n}) - G_{0,f_0}(\beta_{0,f_0})\right\}
\\
&= \frac{1}{n} \sum_{i=1}^n \left\{ \phi_{P_0,f}(Z_i; \tilde{\beta}_{n,f_n}) - \phi_{P_0,f}(Z_i; \beta_{0,f_0}) \right\} + o_P(n^{-1/2})
\\
&\leq \frac{1}{n} \sum_{i=1}^n \sup_{f \in \mathcal{F}, \beta \in \mathcal{B}}\left\{ \phi_{P_0,f}(Z_i; \beta) - \phi_{P_0,f}(Z_i; \beta_{0,f_0}) \right\} + o_P(n^{-1/2})
\\
&= O_P(n^{-1/2}).
\end{align*}
From the above argument, we can conclude that the remainder is at least $O_P(n^{-1/2})$.

Now, we have under the smoothness assumption in \eqref{margin} that
\begin{align*}
\int \left\{\phi _{P_0,f_n}(z; \tilde{\beta}_{n, f_n}) - \phi _{P_0,f_0}(z; \beta_{0, f_0}) \right\}^2dP_0(z) = o_P(1).
\end{align*}
We can now apply lemma 19.24 of \cite{van2000asymptotic} to conclude that
\begin{align*}
\frac{1}{n} \sum_{i=1}^n \left\{ \phi_{P_0,f}(Z_i; \tilde{\beta}_{n,f_n}) - \phi_{P_0,f}(Z_i; \beta_{0,f_0}) \right\} = o_P(n^{-1/2}),
\end{align*}
thereby completing the proof.

\subsubsection*{Proof of Theorem 4}

Let $\mathcal{L}$ be the space of bounded Lipschitz-1 functions $\ell: \mathbb{R} \to [-1,1]$.
That is, any $\ell$ in $\mathcal{L}$ satisfies $|\ell(a_1) -\ell(a_2)| \leq |a_1 - a_2|$ for any $a_1, a_2 \in \mathbb{R}$.
Let $E_\xi$ denote the expectation of a random variable with respect to the distribution of $\xi$ (treating $Z$ as fixed).
We show that
\begin{align*}
\sup_{\ell \in \mathcal{L}}\left| E_\xi \left[ \ell\left(\left\{n T_n^\xi \right\}^{1/2} \right) \right] - E_0\left[\ell\left( \sup_{f \in \mathcal{F}} \left|\left\{2 G''_{0,f}(0)\right\}^{-1/2}\mathbb{H}(f)\right| \right) \right] \right|
\end{align*}
converges to zero in outer probability.
This is equivalent to weak convergence due to Portmanteau lemma \cite[see, e.g. Lemma 18.9 of][]{van2000asymptotic}.

Let $\mathcal{\ell}^\infty(\mathcal{F})$ denote the space of bounded functionals on $\mathcal{F}$, and let $\mathcal{E}$ be the space of Lipschitz-1 functionals  $e:\ell^\infty(\mathcal{F}) \to [-1, 1]$.
That is, for $F_1, F_2$ in $\ell^\infty(\mathcal{F})$, any $e \in \mathcal{E}$ satisfies $|e(F_1) - e(F_2)| \leq \sup_{f \in \mathcal{F}}|F_1(f) - F_2(f)|$.
We now define:
\begin{align*}
&\Lambda^\xi_n: f \mapsto \{2 G''_{\hat{P}_n,f}(0)\}^{-1/2}\left[n^{-1/2}\sum_{i=1}^n \phi'_{\hat{P}_n, f}(Z_i;0) \right],
\\
&\Lambda_0: f \mapsto \{2 G''_{P_0,f}(0)\}^{-1/2} \mathbb{H}(f).
\end{align*}
It is shown by \cite{hudson2021inference} that under Assumptions C1-C3,
\begin{align*}
\sup_{e \in \mathcal{E}} \left| E_\xi\left[e\left(\Lambda_n^\xi\right)\right] - E_0\left[e\left(\Lambda_0\right) \right] \right|.
\end{align*}
The proof is completed by recognizing that for any $\ell \in \mathcal{L}$, the functional $F \mapsto  \left|\ell\left(\sup_{f \in \mathcal{F}}\left|F(f)\right|\right) \right|$ is contained within $\mathcal{E}$.

\subsubsection*{Proof of Theorem 5}

\textbf{Case 1: $\Psi_0 = 0$}

We first consider the setting in which $\Psi_0 = 0$.
Let $\ell$ be a Lipschitz-1 function on $\mathbb{R}$. 
As in the proof of Theorem 4, let $\mathcal{L}$ be the space of Lipschitz-1 functions on $\mathbb{R}$.
We show that 
\begin{align*}
&\sup_{\ell \in \mathcal{L}}\left| E_\xi \left[\ell\left(\left\{ \pi_n n T^\xi_n + (1 - \pi_n) n U^\xi_n \right\}^{1/2}\right)\right] - 
E_0\left[\ell\left( \sup_{f \in \mathcal{F}} \left|\left\{2 G''_{0,f}(0)\right\}^{-1/2}\mathbb{H}(f)\right| \right) \right]  \right|
\end{align*}
converges to zero in outer probability.

First, by applying the triangle inequality and invoking the Lipschitz property, we have
\begin{align*}
&\sup_{\ell \in \mathcal{L}}\left| E_\xi \left[\ell\left(\left\{ \pi_n n T^\xi_n + (1 - \pi_n) n U^\xi_n \right\}^{1/2}\right)\right] - 
E_0\left[\ell\left( \sup_{f \in \mathcal{F}} \left|\left\{2 G''_{0,f}(0)\right\}^{-1/2}\mathbb{H}(f)\right| \right) \right]  \right| \leq A_{n} + B_{n},
\end{align*}
where we define
\begin{align*}
&A_{n} := \left|E_\xi\left[\ell\left(\left|n T^\xi_n\right|^{1/2}\right)\right] -E_0\left[\ell\left( \sup_{f \in \mathcal{F}} \left|\left\{2 G''_{0,f}(0)\right\}^{-1/2}\mathbb{H}(f)\right| \right) \right] \right|,
\\
&B_{n} :=  E_\xi\left|\left\{ (1 - \pi_n) n U^\xi_n + \pi_n n T^\xi_n\right\}^{1/2} - |n T^\xi_n|^{1/2}\right|. 
\end{align*}
We have already shown in Theorem 4 that the first term $A_{n}$ converges to zero in outer probability, so it only remains to verify this for the second term $B_n$.

By the reverse triangle inequality, we have
\begin{align*}
B_{n} \leq (1 - \pi_n) \left\{ E_\xi\left[ \left(n U^\xi_n\right)^{1/2}\right] + E_\xi\left[\left( n T^\xi_n\right)^{1/2} \right] \right\}.
\end{align*}
Because $(1 - \pi_n) = o_P(1)$ when $\Psi_0 = 0$, it suffices to show that $E_\xi\left[ \left(n U_n\right)^{1/2}\right]$ and $E_\xi\left[\left( n T_n\right)^{1/2} \right]$ are both $O_P(1)$.

We first show that $E_\xi\left[ \left(n U_n\right)^{1/2}\right]$ is bounded in probability.
First, we have by Jensen's inequality that
\begin{align*}
E_\xi\left[ \left(n U_n\right)^{1/2}\right] \leq E_\xi\left[ n U_n\right]^{1/2}.
\end{align*}
Now, by Taylor's theorem, we have
\begin{align*}
E_\xi \left[\left|\sum_{i=1}^n \xi_i\left\{\phi_{\hat{P}_n, f_n}(Z_i; 0) - \phi_{\hat{P}_n, f_n}(Z_i; \tilde{\beta}_{n, f_n})\right\} \right| \right] 
\leq 
E_\xi \left[\sup_{f \in \mathcal{F}}\left|\frac{1}{n^{1/2}}\sum_{i=1}^n \xi_i\left\{\phi'_{\hat{P}_n, f}(Z_i; 0)\right\} \right| \right] \sup_{f \in \mathcal{F}}\left|n^{1/2}\bar{\beta}_{n,f}\right|,
\end{align*}
for some $\{\bar{\beta}_{n,f}: f \in \mathcal{F}\}$ that satisfies $|\bar{\beta}_{n,f}| \leq |\tilde{\beta}_{n,f}|$ for all $f \in \mathcal{F}$.
Because $\sup_{f \in \mathcal{F}}|n^{1/2}\bar{\beta}_{n,f}| = O_P(1)$ under the conditions of Theorem 2, it suffices to show that
\begin{align}
E_\xi \left[\sup_{f \in \mathcal{F}}\left|\frac{1}{n^{1/2}}\sum_{i=1}^n \xi_i\left\{\phi'_{\hat{P}_n, f}(Z_i; 0)\right\} \right| \right] = O_P(1).
\label{RademacherOP1}
\end{align}
By the triangle inequality, we have the upper bound
\begin{align*}
&E_\xi \left[\sup_{f \in \mathcal{F}}\left|\frac{1}{n^{1/2}}\sum_{i=1}^n \xi_i\left\{\phi'_{\hat{P}_n, f}(Z_i; 0)\right\} \right| \right]  \leq B_{n,1} + B_{n,2} + B_{n,3},
\end{align*}
where we define
\begin{align*}
&B_{n,1} = E_\xi \left[\sup_{f \in \mathcal{F}}\left|\frac{1}{n^{1/2}}\sum_{i=1}^n \xi_i \int \left\{\phi'_{\hat{P}_n, f}(z; 0) - \phi'_{P_0,f}(z)\right\} dP_0(z) \right| \right],
\\
&B_{n,2} = E_\xi \left[\sup_{f \in \mathcal{F}}\left|\frac{1}{n^{1/2}}\sum_{i=1}^n \xi_i\left[\left\{\phi'_{\hat{P}_n, f}(Z_i; 0) - \phi'_{P_0, f}(Z_i; 0)  \right\} - \int \left\{\phi'_{\hat{P}_n, f}(z; 0) - \phi'_{P_0,f}(z)\right\} \right] dP_0(z) \right| \right],
\\
&B_{n,3} = E_\xi \left[\sup_{f \in \mathcal{F}}\left|\frac{1}{n^{1/2}}\sum_{i=1}^n \xi_i\left\{\phi'_{\hat{P}_n, f}(Z_i; 0)\right\} \right| \right] .
\end{align*}
It can be seen through an application of the Cauchy-Schwarz inequality $B_{n,1} = o_P(1)$ under Assumption C2.
To see that $B_{n,2}$ is bounded in probability, we first note that Assumptions C1 implies that with probability tending to one,
\begin{align*}
B_{n,2} \leq E_{\xi}\left[\sup_{\varphi \in \Phi_\delta} \left|\frac{1}{n^{1/2}} \sum_{i=1}^n \xi_i\varphi(Z_i) \right| \right].
\end{align*}
In view of Markov's inequality, it is sufficient to show that this upper bound has finite expectation.
Lemma 2.3.6 of \cite{van1996weak} implies that
\begin{align*}
E_0 E_{\xi}\left[\sup_{\varphi \in \Phi_\delta} \left|\frac{1}{n^{1/2}} \sum_{i=1}^n \xi_i\varphi(Z_i) \right| \right] \leq 2 E_0\left[\sup_{\varphi \in \Phi_\delta} \left|\frac{1}{n^{1/2}} \sum_{i=1}^n \varphi(Z_i) \right| \right].
\end{align*}
Because under Assumption C3, $\Phi_\delta$ has finite bracketing integral, we have by Corollary 19.35 of \cite{van2000asymptotic} that
\begin{align*}
\left[\sup_{\varphi \in \Phi_\delta} \left|\frac{1}{n^{1/2}} \sum_{i=1}^n \varphi(Z_i) \right| \right] < \infty,
\end{align*}
thereby establishing that $B_{n,2} = O_P(1)$.
That $B_{n,3} = O_P(1)$ follows from a similar argument.

To argue that  $E_\xi\left[\left( n T^\xi_n\right)^{1/2} \right]$ is $O_P(1)$, we use the same argument as is used to show that \eqref{RademacherOP1} holds.
In brief, the result follows from the facts that ($\mathrm{i}$) both $\left\{G''_{\hat{P},f}: f \in \mathcal{F}\right\}$ and $\{z \mapsto \phi'_{\hat{P}_n,f}(z;0): f\in \mathcal{F}\}$ are uniformly consistent under Assumption C2, and ($\mathrm{ii}$) the class 
\begin{align*}
\bar{\Phi}'_{\delta_2} := \{ z \mapsto \left[G''_{P,f}(0)\right]^{-1}\phi'_{P,f}(z; 0) - \left[G''_{P_0,f}(0)\right]^{-1}\phi'_{P_0,f}(z;0): f \in \mathcal{F}, \|Q_P - Q_{P_0} \|_{\mathcal{Q}} \leq \delta_2\}
\end{align*}
is $P_0$-Donsker with finite bracketing integral under Assumption C3.
\\ \\
\textbf{Case 2: $\Psi_0 > 0$}

We now consider the setting where $\Psi_0 > 0$.
We show that
\begin{align*}
&\sup_{\ell \in \mathcal{L}}\left| E_\xi \left[\ell\left( \pi_n n^{1/2} T^\xi_n + (1 - \pi_n) n^{1/2} U^\xi_n \right)\right] - 
E_0\left[ \ell\left(|\mathbb{I} |\right) \right] \right|,
\end{align*}
converges to zero in outer probability.
Similarly as for Case 1, we have by the triangle inequality that
\begin{align*}
&\sup_{\ell \in \mathcal{L}}\left| E_\xi \left[\ell\left( \pi_n n^{1/2} T^\xi_n + (1 - \pi_n) n^{1/2} U^\xi_n \right)\right] - 
E_0\left[ \ell\left(|\mathbb{I}|\right) \right] \right| \leq  A_n + B_n,
\end{align*}
where we define
\begin{align*}
&A_n := \sup_{\ell \in \mathcal{L}}\left| E_\xi \left[\ell\left(n^{1/2} U^\xi_n \right)\right] - E_0[\ell(\mathbb{I})] \right|,
\\
&B_n :=\pi_n \left\{ E_\xi\left[n^{1/2} U_n^\xi \right] + E_\xi\left[n^{1/2} T_n^\xi\right] \right\}.
\end{align*}

We first argue that $A_n$ converges to zero in outer probability. 
First, we have under assumption that the function
\begin{align*}
z \mapsto \phi_{\hat{P}_n, f_n}(z; 0) - \phi_{\hat{P}_n, f_n}(z; \tilde{\beta}_{n, f_n}) 
\end{align*}
is contained within a $P_0$-Donsker class with probability tending to one.
\sloppy Also we have under Assumption C2 that $\int \left\{\phi_{\hat{P}_n,f}(z; 0) - \phi_{P_0,f}(z;0) \right\}^2dP_0(z) = o_P(1)$.
 We now argue that  $\int \left\{\phi_{\hat{P}_n,f}(z;\tilde{\beta}_{n,f_n}) - \phi_{P_0, f}(z; \tilde{\beta}_{0,f_0}) \right\}^2dP_0(z)  = o_P(1)$.
We have the upper bound
\begin{align*}
&\int \left\{\phi_{\hat{P}_n,f}(z;\tilde{\beta}_{n,f_n}) - \phi_{P_0, f}(z; \tilde{\beta}_{0,f_0}) \right\}^2dP_0(z) 
\leq 
\\
&2\left[\int \left\{\phi_{\hat{P}_n,f}(z;\tilde{\beta}_{n,f_n}) - \phi_{P_0, f}(z; \tilde{\beta}_{n,f_n}) \right\}^2dP_0(z)
+
\int \left\{\phi_{P_0,f_n}(z;\tilde{\beta}_{n,f_n}) - \phi_{P_0, f}(z; \beta_{0,f_0}) \right\}^2dP_0(z)\right].
\end{align*}
Under Assumption C2, we have
\begin{align*}
\int \left\{\phi_{\hat{P}_n,f}(z;\tilde{\beta}_{n,f_n}) - \phi_{P_0, f}(z; \tilde{\beta}_{n,f_n}) \right\}^2dP_0(z) 
\leq
\sup_{f \in \mathcal{F}, \beta \in \mathcal{B}}\int \left\{\phi_{\hat{P}_n,f}(z;\beta) - \phi_{P_0, f}(z; \beta) \right\}^2dP_0(z) = o_P(1).
\end{align*}
Additionaly, we have 
\begin{align*}
\int \left\{\phi_{P_0,f_n}(z;\tilde{\beta}_{n,f_n}) - \phi_{P_0, f}(z; \beta_{0,f_0}) \right\}^2dP_0(z) = o_P(1)
\end{align*}
under the conditions of Theorem 3.
Now, by Theorem 2 of \cite{hudson2021inference}, we can conclude that $|A_n|$ converges to zero in outer probability.

We now argue that $B_n = o_P(1)$.
Because $\pi_n = o_P(1)$, we only need to show that $E_\xi\left[n^{1/2} U_n^\xi \right] $ and $E_\xi\left[n^{1/2} T_n^\xi \right]$ are bounded in probability.
That $E_\xi\left[n^{1/2} U_n^\xi \right] = O_P(1)$ follows from the same argument as was used to show that \eqref{RademacherOP1} holds in Case 1.

To argue that $E_\xi\left[n^{1/2} T_n^\xi \right] = O_P(1)$, we begin by applying the Cauchy-Schwarz inequality and invoking Assumption C2 to get
\begin{align*}
\left\{T_n^\xi\right\}^{1/2} &\leq \left[ \frac{1}{n}\sum_{i=1}^n \xi_i^2 \right]^{1/2}\left[ \sup_{f \in \mathcal{F}}\frac{1}{n}\sum_{i=1}^n \left\{ [G''_{\hat{P}_n,f}(0)]^{-1} \phi'_{\hat{P}_{n},f}(Z_i; 0)\right\}^2 \right]^{1/2}  
\\
&= \left[ \sup_{f \in \mathcal{F}}\frac{1}{n}\sum_{i=1}^n \left\{ [G''_{\hat{P}_n,f}(0)]^{-1} \phi'_{\hat{P}_{n},f}(Z_i; 0)\right\}^2 \right]^{1/2}.
\end{align*}
Now, by the triangle inequality,
\begin{align*}
&\left[ \sup_{f \in \mathcal{F}}\frac{1}{n}\sum_{i=1}^n \left\{ [G''_{\hat{P}_n,f}(0)]^{-1} \phi'_{\hat{P}_{n},f}(Z_i; 0)\right\}^2 \right]^{1/2} 
\leq
\\
&\left[ \sup_{f \in \mathcal{F}}\frac{1}{n}\sum_{i=1}^n \left\{ [G''_{P_0,f}(0)]^{-1} \phi'_{P_0,f}(Z_i; 0)\right\}^2 \right]^{1/2} 
+
\\
&\left[ \sup_{f \in \mathcal{F}}\frac{1}{n}\sum_{i=1}^n \left\{[G''_{\hat{P}_n,f}(0)]^{-1}\phi'_{\hat{P}_{n},f}(Z_i; 0) - [G''_{P_0,f}(0)]^{-1}\phi_{P_0}(Z_i; 0)\right\}^2 \right]^{1/2}.
\end{align*}
Because $\{[G''_{0,f}(0)]^{-1}\phi'_{0,f}(\cdot;0): f \in \mathcal{F}\}$ is a $P_0$-Donsker class with finite squared envelope function, we have by Lemma 2.10.4 of \cite{van1996weak} that $\left\{ \left[[G''_{0,f}(0)]^{-1}\phi'_{0,f}(\cdot;0) \right]^2: f \in \mathcal{F}\right\}$ is a $P_0$-Glivenko-Cantelli class, and so
\begin{align*}
\left[ \sup_{f \in \mathcal{F}}\frac{1}{n}\sum_{i=1}^n \left\{ [G''_{P_0,f}(0)]^{-1} \phi'_{P_{0},f}(Z_i; 0)\right\}^2 \right]^{1/2} = O_P(1).
\end{align*}
Now, by the triangle inequality
\begin{align*}
\sup_{f \in \mathcal{F}}&\left[\frac{1}{n}\sum_{i=1}^n \left\{ [G''_{\hat{P}_n,f}(0)]^{-1} \phi'_{\hat{P}_{n},f}(Z_i; 0) - [G''_{P_0,f}(0)]^{-1} \phi_{P_0}(Z_i; 0)\right\}^2 \right]^{1/2} \leq
\\
\sup_{f \in \mathcal{F}}\Bigg[&\frac{1}{n}\sum_{i=1}^n \left\{ [G''_{\hat{P}_n,f}(0)]^{-1} \phi'_{\hat{P}_{n},f}(Z_i; 0) - [G''_{P_0,f}(0)]^{-1}\phi_{P_0}(Z_i; 0)\right\}^2 - 
\\
&\int \left\{ [G''_{\hat{P}_n,f}(0)]^{-1} \phi'_{\hat{P}_n,f}(z;0) - [G''_{P_0,f}(0)]^{-1} \phi'_{P_0, f}(z;0) \right\}^2 dP_0 \Bigg]^{1/2} + 
\\
\sup_{f\in\mathcal{F}}&\left[\int \left\{[G''_{\hat{P}_n,f}(0)]^{-1}\phi'_{\hat{P}_n,f}(z;0) - \int [G''_{P_0,f}(0)]^{-1}\phi'_{P_0, f}(z;0) \right\}^2 dP_0 \right]^{1/2}.
\end{align*}
We have by assumption that 
\begin{align*}
\left[\int \left\{[G''_{\hat{P}_n,f}(0)]^{-1}\phi'_{\hat{P}_n,f}(z;0) - \int [G''_{P_0,f}(0)]^{-1}\phi'_{P_0, f}(z;0) \right\}^2 dP_0 \right]^{1/2}  = o_P(1).
\end{align*}
Additionally, Assumption C2 and Lemma 2.10.4 of \citep{van1996weak} imply that the class \sloppy $\left\{\left\{[G''_{P,f}(0)]^{-1}\phi'_{P,f}(\cdot;0) - [G''_{0,f}(0)]^{-1}\phi'_{P_0,f}(\cdot)\right\}^{2} \right\}$ is $P_0$-Glivenko-Cantelli with probability tending to one.
Therefore,
\begin{align*}
&\Bigg[ \sup_{f \in \mathcal{F}}\frac{1}{n}\sum_{i=1}^n \left\{ [G''_{\hat{P}_n,f}(0)]^{-1} \phi'_{\hat{P}_{n},f}(Z_i; 0) - [G''_{P_0,f}(0)]^{-1}\phi_{P_0}(Z_i; 0)\right\}^2 - 
\\
&\int \left\{ [G''_{\hat{P}_n,f}(0)]^{-1} \phi'_{\hat{P}_n,f}(z;0) - [G''_{P_0,f}(0)]^{-1} \phi'_{P_0, f}(z;0) \right\}^2 dP_0 \Bigg]^{1/2} = o_P(1).
\end{align*}
Thus, we have that $\left\{T_{n}^\xi\right\}^{1/2} \leq O_P(1)$.
This allows us to write
\begin{align*}
E_\xi\left[n^{1/2} T_n^\xi \right] \leq O_P(1) E_\xi\left[\left\{n T_n^\xi\right\}^{1/2} \right].
\end{align*}
That  $E_\xi\left[\left\{n T_n^\xi\right\}^{1/2}\right] = O_P(1)$ follows from the argument presented in Case 1.
This completes the proof.

\subsubsection*{Proof of Lemma 1}

Suppose a given distribution $P$ in $\mathcal{M}$ has density $p$ with respect to a dominating measure $\nu$,
and let $\eta: \mathcal{Z} \to \mathbb{R}$ be a fixed function that has mean zero and finite variance under $P$.
Let $P_\epsilon$ be a one-dimensional parametric sub-model for $P$ indexed by the parameter $\epsilon$, which satisfies the following:
\begin{enumerate}
\item The sub-model passes through $P$ at $\epsilon = 0$ -- that is, $P_{\epsilon} = P$ at $\epsilon = 0$
\item The density of the parametric sub-model is given by $p_\epsilon$, and the score function is given by $\eta$ at $\epsilon = 0$. That is,
\begin{align*}
\frac{d}{d\epsilon} \log p_{\epsilon}(z) \bigg\vert_{\epsilon = 0} = \eta(z).
\end{align*}
\end{enumerate}
We refer to $\frac{d}{d\epsilon} G_{P_\epsilon,f}(\beta) \vert_{\epsilon = 0}$ as the \textit{pathwise derivative} of $G_{P_\epsilon,f}(\beta)$.
The nonparametric efficient influence function $\phi_{P,f}(\cdot; \beta)$  is the unique function that satisfies the following two properties:
\begin{enumerate}
\item For every $\eta$, $\frac{d}{d\epsilon} G_{P_\epsilon}(\beta) = \int \phi_{P,f}(z; \beta) \eta(z) p(z) d\nu(z)$.
\item $\phi_{P,f}(Z; \beta)$ has mean zero under $P$. That is, $\int \phi_{P,f}(z;\beta) dP(z) = 0$.
\end{enumerate}
We can therefore find the efficient influence function by calculating the pathwise derivative.

Let $P_\epsilon$ have density
\begin{align*}
p_\epsilon = p(1 + \epsilon \eta).
\end{align*}
We can approximate any density in a neighborhood of $p$ with such a sub-model.
Any distribution in a small neighborhood of $P$ can be approximated using a sub-model of this form.

The goodness-of-fit under $P_\epsilon$ is given by
\begin{align*}
G_{P_\epsilon,f}(\beta) = \int \left\{y - \mu_{P_\epsilon,Y}(w) - \beta f(w,x) \right\}^2p(z)\{1 + \epsilon \eta (z)\}  d\nu(z).
\end{align*}
Through a simple calculation, it can be shown that
\begin{align*}
\frac{d}{d\epsilon} \mu_{P_\epsilon,Y}(w) \bigg|_{\epsilon = 0} = \int \{y_1 - \mu_{Y,P}(w)\}\eta(w,x_1,y_1) p(x_1,y_1|w)\nu(dx_1, dy_1),
\end{align*}
where $p(\cdot|w)$ denotes the conditional density of $(X,Y)$ given that $W = w$, under $P$. We now have
\begin{align*}
\frac{d}{d\epsilon} G_{P_\epsilon}(\beta) \bigg\vert_{\epsilon = 0} = &\int \left\{y - \mu_{P,Y}(w) - \beta f(w,x) \right\}^2p(z)\eta(z)  d\nu -
\\
&2\int\{y - \mu_{P,Y}(w) - \beta f(w,x)\}\left\{\int \{y_1 - \mu_{P,Y}(w)\}\eta(w,x_1,y_1) p(x_1,y_1|w)\nu(dx_1, dy_1)\right\}p(z) d\nu(z),
\\ \\
= &\int \left\{y - \mu_{P,Y}(w) - \beta f(w,x) \right\}^2p(z)\eta(z)  d\nu +
\\
&2\int\left[
\int  \beta f(w,x_1)p(x_1,y_1|w)\nu(dx_1,dy_1)\right] \{y - \mu_{P,Y}(w)\}\eta(w,x,y) p(z) d\nu 
\\
\\
= & \int \left[\left\{y - \mu_{P,Y}(w) - \beta f(w,x)\right\}^2 + 2 \beta \mu_{P,f}(w,x)\left\{y - \mu_{P,Y}(w)\right\} \right] \eta(w,x,y)p(z) d\nu,
\end{align*}
where the second equality follows from an application of the law of total expectation to the second summand.
The ``non-mean-centered'' efficient influence function is thus given by
\begin{align*}
z = (w,x,y) \mapsto \left\{y - \mu_{Y,P}(w) - \beta f(w,x)\right\}^2 + 2 \beta \mu_{f,P}(w)\left\{y - \mu_{Y,P}(w)\right\}.
\end{align*}
The result is completed by centering the above function about its mean.

\subsubsection*{Proof of Lemma 2}

We write the estimation error for the one-step estimator as
\begin{align*}
\tilde{G}_{n,f}(\beta) - G_{0,f}(\beta) = \frac{1}{n}\sum_{i=1}^n \phi_{P_0,f}(Z_i; \beta) + R^{\mathrm{i}}_{n,f}(\beta) + R^{\mathrm{ii}}_{n,f}(\beta),
\end{align*}
where the remainder terms are
\begin{align*}
&R_{n,f}^{\mathrm{i}}(\beta) = \sum_{i=1}^n \left\{\phi_{n,f}(Z_i;\beta) - \phi_{P_0,f}(Z_i;\beta) \right\} - \int \left\{ \phi_{n,f}(Z_i;\beta) - \phi_{P_0,f}(Z_i;\beta) \right\} dP_0(z),
\\
&R_{n,f}^{\mathrm{ii}}(\beta) = \int \phi_{n,f}(z; \beta) dP_0(z) + \left\{ G_{n,f}(\beta) - G_{P_0,f}(\beta) \right\} .
\end{align*}
Following from our discussion in Section \ref{sec:submodel-inference}, it suffices to argue each of the following:
\begin{align*}
&\sup_{f \in \mathcal{F}, \beta \in \mathcal{B}} |R^{\mathrm{i}}_{n,f}(\beta)| = o_P(n^{-1/2}),\quad
 \sup_{f \in \mathcal{F}} \left|\frac{d}{d\beta}\left\{R^{\mathrm{i}}_{n,f}(\beta)\right\}_{\beta= 0}\right| = o_P(n^{-1/2}),\quad
 \sup_{f \in \mathcal{F}} \left|\frac{d^2}{d\beta^2}\left\{R^{\mathrm{i}}_{n,f}(\beta)\right\}_{\beta= 0}\right| = o_P(1),
 \\
 &\sup_{f \in \mathcal{F}, \beta \in \mathcal{B}} |R^{\mathrm{ii}}_{n,f}(\beta)| = o_P(n^{-1/2}),\quad
 \sup_{f \in \mathcal{F}} \left|\frac{d}{d\beta}\left\{R^{\mathrm{ii}}_{n,f}(\beta)\right\}_{\beta= 0}\right| = o_P(n^{-1/2}),\quad
 \sup_{f \in \mathcal{F}} \left|\frac{d^2}{d\beta^2}\left\{R^{\mathrm{ii}}_{n,f}(\beta)\right\}_{\beta= 0}\right| = o_P(1).
\end{align*}

First, we argue that $\sup_{f \in \mathcal{F}, \beta \in \mathcal{B}} |R_{n,f}(\beta)| = o_P(n^{-1/2})$.
It is shown in the proof of Lemma 19.26 of van der Vaart that this convergence rate is achieved when 
\begin{align*}
\sup_{f \in \mathcal{F}, \beta \in \mathcal{B}} \int \left\{\phi_{n,f}(z; \beta) - \phi_{P_0,f}(z; \beta) \right\}^2dP_0(z) = o_P(1),
\end{align*}
and when the class $\{\phi_{n,f}(\cdot;\beta) - \phi_{P_0,f}(\cdot; \beta): f \in \mathcal{F}, \beta \in \mathcal{B} \}$ is contained within a $P_0$-Donsker class probability tending to one.
That the influence functions are uniformly consistent follows a consequence of the rate conditions on the nuisance parameter estimators, and the Donsker condition holds by assumption.
Similarly, that $\sup_{f \in \mathcal{F}}|\frac{d}{d\beta}\{R_{n,f(\beta)}\}_{\beta = 0}| = o_P(n^{-1/2})$ and $\sup_{f \in \mathcal{F}}|\frac{d^2}{d\beta^2}\{R_{n,f(\beta)}\}_{\beta = 0}| = o_P(n^{-1/2})$ follow from consistency of nuisance estimators and the assumed complexity constraints.

Now, we argue that $\sup_{f \in \mathcal{F}, \beta \in \mathcal{B}}\left|R^{\mathrm{ii}}_{n,f}(\beta)\right| = o_P(n^{-1/2})$. This remainder term has the exact representation
\begin{align*}
&\int \phi_{n,f}(z; \beta) dP_0(z) + \left\{ G_{n,f}(\beta) - G_{P_0,f}(\beta) \right\} =
\\
&\int \left\{y - \mu_{n,Y}(w) - \beta f(w,x)\right\}^2 + 2\beta\left\{y - \mu_{n,Y}(w)\right\}\mu_{n,f}(w) -  \left\{y - \mu_{0,Y}(w) - \beta f(w,x)\right\}^2 dP_0(z) =
\\ \\
&\int \left[ \left\{y - \mu_{n, Y}(w)\right\} - \left\{y - \mu_{P_0,Y}(w)\right\} \right]
\left[ \left\{y - \mu_{n,Y}(w)\right\} + \left\{y - \mu_{P_0,Y}(w)\right\} \right]dP_0(z) - 
\\
&\int 2\beta\left[\left\{\mu_{P_0, Y}(w) - \mu_{n,Y}(w)\right\}f(w,x) - \left\{y - \mu_{n,Y}(w)\right\}\mu_{n,f}(w) \right]dP_0(z) =
\\ \\
&\int \left\{\mu_{P_0,Y}(w) - \mu_{n,Y}(w)\right\}^2dP_0(z) +  2\beta\left\{\mu_{P_0,Y}(w) - \mu_{n,Y}(w)\right\}\left\{ \mu_{f,P_0}(w) - \mu_{f,P}(w) \right\}   dP_0(z).
\end{align*}
It can seen that $\sup_{f \in \mathcal{F}, \beta \in \mathcal{B}}\left|R^{\mathrm{ii}}_{n,f}(\beta)\right| = o_P(n^{-1/2})$ when the rate conditions on the nuisance estimators are met.
The derivative of the second remainder term is
\begin{align*}
\frac{d}{d\beta} R^{\mathrm{ii}}_{n,f}(\beta) \bigg\vert_{\beta = 0} =  \int   2\left\{\mu_{P_0,Y}(w) - \mu_{n,Y}(w)\right\}\left\{ \mu_{f,P_0}(w) - \mu_{f,P}(w) \right\}   dP_0(z),
\end{align*}
and so $\sup_{f \in \mathcal{F}}|\frac{d}{d\beta}\{R^{\mathrm{ii}}_{n,f(\beta)}\}_{\beta = 0}| = o_P(n^{-1/2})$ under the rate conditions as well.
Finally, it is easily seen that $\sup_{f \in \mathcal{F}}|\frac{d^2}{d\beta^2}\{R^{\mathrm{ii}}_{n,f(\beta)}\}_{\beta = 0}| = 0$.

\end{document}